\documentclass[fleqn,usenatbib]{mnras}

\usepackage{newtxtext,newtxmath}

\usepackage[T1]{fontenc}
\usepackage[normalem]{ulem}

\DeclareRobustCommand{\VAN}[3]{#2}
\let\VANthebibliography\thebibliography
\def\thebibliography{\DeclareRobustCommand{\VAN}[3]{##3}\VANthebibliography}

\usepackage{graphicx}	
\usepackage{amsmath}	
\usepackage{multirow}
\usepackage[dvipsnames]{xcolor}

\title[Modelling AGN in galaxy clusters]{Scrutinising evidence for the triggering of Active Galactic Nuclei in the outskirts of massive galaxy clusters at $z\approx1$}
\author[I. Mu\~noz-Rodr\'iguez et al.]{
Iv\'an Mu\~noz Rodr\'iguez$^{1,2}$\thanks{E-mail: ivan.rodriguez@noa.gr},
Antonis Georgakakis$^{1}$,
Francesco Shankar$^{2}$,
\'Angel Ruiz$^{1}$,
Silvia Bonoli$^{3,4}$
\newauthor Johan Comparat$^5$,
Elias Koulouridis$^{1}$,
Andrea Lapi$^{6}$,
Cristina Ramos Almeida$^{7,8}$
\\
$^{1}$Institute for Astronomy and Astrophysics, National Observatory of Athens, V. Paulou \& I. Metaxa 11532, Greece
\\
$^{2}$School of Physics and Astronomy, University of Southampton, Highfield SO17 1BJ, Southampton, UK\\
$^{3}$Donostia International Physics Center (DIPC), Manuel Lardizabal Ibilbidea, 4, San Sebastián, Spain\\
$^{4}$Ikerbasque, Basque Foundation for Science, E-48013 Bilbao, Spain\\
$^{5}$Max-Planck-Institut für Extraterrestrische Physik (MPE), Giessenbachstrasse 1, 85748 Garching bei München, Germany\\
$^{6}$SISSA, Via Bonomea 265, 34136 Trieste, Italy\\
$^7$Instituto de Astrofísica de Canarias, Calle Vía Láctea, s/n, 38205 La Laguna, Tenerife, Spain\\
$^8$Departamento de Astrofísica, Universidad de La Laguna, 38206 La Laguna, Tenerife, Spain
}

\date{Accepted XXX. Received YYY; in original form ZZZ}

\pubyear{2022}

\begin{document}
\label{firstpage}
\pagerange{\pageref{firstpage}--\pageref{lastpage}}
\maketitle
\begin{abstract}
Environmental effects are believed to play an important yet poorly understood role in triggering accretion events onto the supermassive black holes (SMBHs) of galaxies (Active Galactic Nuclei; AGN). Massive clusters, which represent the densest structures in the Universe, provide an excellent laboratory to isolate environmental effects and study their impact on black hole growth. In this work, we critically review observational evidence for the preferential activation of SMBHs in the outskirts of galaxy clusters. We develop a semi-empirical model under the assumption that the incidence of AGN in galaxies is independent of environment. We demonstrate that the model is broadly consistent with recent observations on the AGN halo occupation at $z=0.2$, although it may overpredict satellite AGN in massive halos at that low redshift. We then use this model to interpret the projected radial distribution of X-ray sources around high redshift ($z\approx1$) massive ($>5 \times 10^{14} \, M_\odot$) clusters, which show excess counts outside their virial radius. Such an excess naturally arises in our model as a result of sample variance. Up to 20\% of the simulated projected radial distributions show excess counts similar to the observations, which are however, because of background/foreground AGN and hence, not physically associated with the cluster. Our analysis emphasises the importance of projection effects and shows that current observations of $z\approx1$ clusters remain inconclusive on the activation of SMBHs during infall.
\end{abstract}




\section{Introduction}\label{sec:intro}
A major challenge in current astrophysical research is to understand the formation and evolution of galaxies in the Universe. The difficulty in addressing this issue is that the relevant physical processes, such as the cooling of gas, the formation of stars and the injection of energy and metals into the interstellar medium by e.g. dying stars, are complex, interconnected and operate over a wide range of temporal and spatial scales \citep[e.g.][]{Benson2010, Somerville_Dave2015}. A development that has changed the way we view galaxy evolution has been the realisation that nearly every spheroidal galaxy hosts at its nuclear regions a black hole with a mass that may exceed $10^9 \,\rm{M}_\odot$ that appears to correlate with the mass of the stellar population \citep[e.g.][]{Graham2011, Kormendy_Ho2013}. Although these supermassive black holes (SMBHs) are gravitationally insignificant for their galaxies, theoretical arguments and observational results suggest that their energy output during their growth phases has a strong impact on the interstellar medium and can affect the evolution path of their hosts \citep[e.g.][]{Fabian2012}. Understanding in detail this symbiotic relationship is therefore important for painting a complete picture of galaxy evolution. A first step toward this goal is to understand the physical conditions that produce accretion flows onto SMBHs, thereby leading to their growth and the release of energy that is observed as Active Galactic Nuclei (AGN).

The activation of SMBHs relies on two factors. The availability of cold gas in the galaxy to fuel these compact objects and a mechanism that is able to drive this material to the galactic centre in the vicinity of the SMBH. Secular processes that occur during the lifetime of galaxies can generate conditions that fulfil the requirements above and hence promote the growth of black holes. For example, recycled gas produced by normal stellar evolution can provide sufficient reservoirs of available fuel and lead to recursive cycles of black hole accretion flows \citep[e.g.][]{Ciotti2007}.  Disk instabilities \citep[e.g.][]{Hopkins_Hernquist2006, Gatti2016} and galactic bars \citep{Cisternas2015} are efficient in removing angular momentum from the interstellar gas thereby driving it toward the central regions of the galaxy where it can be accreted by the SMBH. In the early Universe, the direct collapse of low angular momentum gaseous baryons is proposed to lead to starburst events as well as the rapid growth of  black holes and  ultimately produce the progenitors of present-day early-type massive galaxies \citep{Shi2017, Lapi2018}. In addition to the in-situ processes above, environmental effects are also thought to play an important role in modulating accretion flows onto SMBHs. For example, in the case of galaxies in dense regions of the cosmic web, ram pressure may initially compress the cold gas in the nuclear regions of galaxies \cite[][]{Marshall2018, Ricarte2020} and hence, promote accretion flows onto SMBHs \cite[][]{peluso22}. In the longer term, however, this process acts to deplete the cold gas reservoirs of the galaxies \cite[][]{Steinhauser2016} thereby suppressing the growth of their black holes. Gravitational interactions and mergers are long thought to represent an important AGN trigger \citep[e.g.][]{DiMatteo2005, 2006Koulouridis,2013Koulouridis,Gatti2016} and perhaps the dominant mechanism in the case of the most luminous SMBH accretion events \cite[e.g.][]{Glikman2015, Araujo2023}. At high redshift flows of cold gas from the cosmic web onto galaxies are proposed to be common leading to both intense star-formation and AGN activity \citep{Bournaud2012, DeGraf2017}. 

This paper focuses on AGN triggering mechanisms that are pertinent to the densest structures in the Universe,  massive galaxy clusters. These systems offer a perfect laboratory for isolating environmental effects to explore how they modulate black hole growth. Additionally, by scanning clusters of galaxies from beyond their outskirts to their cores it is possible to sample a broad range of densities and therefore witness the onset of environmental effects and study their impact as a function of local density. Observations of the fraction of AGN in low redshift ($z\la0.3$) clusters indicate that the nuclear activity in galaxies is suppressed in these dense environments, particularly close to the centre of the potential well \citep{haines12,sabater12,martini13,koulouridis14,Mishra20}. Nonetheless, this trend seems to level off at intermediate redshift \citep[$z\approx0.7$,][]{eastman07,Galametz09,martini09,martini13,Ehlert14,bufanda17} and perhaps reverse at $z\ga$1 \citep[see][]{lehmer09,digby-north10,Krishnan17,Tozzi_2022,monson2023,munozRodriguez2023, Toba2024}. These studies show that the incidence of AGN in clusters of galaxies is similar to the field expectation at $z\approx0.3-0.8$ and exceeds this value at earlier cosmic times. The cluster environment, therefore, appears to promote black hole growth outside the local Universe. Efficient activation of AGN in dense regions points to physical mechanisms that operate preferentially in these environments, such as a higher galaxy interaction rate \citep{Gatti2016} or ram-pressure \citep[][]{Poggianti2017,peluso22}. These processes are expected to be more efficient at the outskirts of clusters \citep[e.g.][]{Toba2024} where the local density is lower and the relative velocities of galaxies smaller.

An infalling population of active black holes may imprint observable features on the radial distribution of AGN within a cluster. There is indeed evidence that the fraction of AGN relative to galaxies is decreasing toward the cluster centre suggesting a higher incidence of AGN at the cluster outskirts \citep{martini09,Ehlert14,dasouza16,Lopes17,Mishra20,stroe21,2024Koulouridis}. Additionally, there are claims that the projected counts of AGN show an excess outside the virial radius of clusters \citep{Johnson03,ruderman05,fassbender12,koulouridis2019}, which could be interpreted as direct evidence of SMBH activation during infall. However, these results remain controversial with a number of studies failing to observe such projected overdensities \citep[][]{Ehlert14,Mo18,Mishra20}. Part of the discrepancy can be attributed to differences in cluster halo masses or cluster dynamical states among the various samples  \citep[e.g.][]{Hashiguchi2023}, AGN selection effects such as flux limits or selection wavelength as well as cluster to cluster variations \citep[see][]{martini07}.

In this work, we revisit claims for an excess of AGN activity in the outskirts of clusters by developing a semi-empirical model to interpret the observed X-ray AGN radial distributions presented by \cite{koulouridis2019}. This work uses {\it Chandra} observations of a well-defined sample of clusters with carefully measured masses and sizes to find a statistically significant excess of X-ray point sources at a distance of about $2.5\,R_{500}$ from the cluster centre, where $R_{500}$ is the radius that encloses a volume with mass density 500 times the critical one of the Universe at the redshift of interest. The  \cite{koulouridis2019} work has a number of key features that greatly facilitate the modelling and interpretation. The first is the transparent selection of the clusters and the corresponding AGN which can be replicated in the modelling. The second is the fact that the radial distributions are expressed in units of $R_{500}$ thereby allowing direct comparison of clusters with different masses and extents. The semi-empirical modelling approach we develop in this paper provides an excellent handle on systematics and selection effects and enables us to explore the impact of projection effects and sample variance in the radial distributions of AGN in clusters of galaxies. Our modelling is based on observationally derived relations to populate dark matter halos extracted from N-body simulations with AGN and galaxies under the assumption that the incidence of AGN does not depend on environment. The comparison with the observations follows the principles of forward modelling to generate realistic cluster observations that include selection effects such as flux limits and the finite {\it Chandra} field of view. Section~\ref{sec:obs} presents the observations and the cluster sample used in this work. Section~\ref{sec:methods} describes the generation of the mock catalogues and the implementation of the different selection effects into the simulations. The comparison of the semi-empirical model predictions with the observations is presented in Section~\ref{sec:results}. Finally, Section~\ref{sec:discussion} discusses the results in the context of the current debate on AGN radial distribution in clusters. We adopt a flat $\Lambda$CDM cosmology with parameters $\Omega_m$ = 0.307, $\Omega_\Lambda$= 0.693, $h$ = 0.678 consistent with the Planck results \citep{planck16}.

\section{Observations}\label{sec:obs}
This work uses {\it Chandra} X-ray observations of massive clusters of galaxies at $z\approx1$ presented by \cite{koulouridis2019}. Their sample is selected using the Sunyaev-Zeldovich effect (SZ) and it is composed by the 5 most massive clusters ($M_{500}^{SZ}\gtrsim5\times10^{14}\rm{M}_\odot$) in the South Pole Telescope and {\it Planck} catalogues at that redshift \citep[see][]{2015Bleem,planck16}. These are the only clusters at that redshift for which detailed analysis of their X-ray profiles have been carried out \citep{Bartalucci2017} to provide robust constraints on their $R_{500}$ ($0.7-1$\,Mpc) and M$_{500}$ (mass range $3 - 8 \times10^{14}\,M_\odot$). \cite{koulouridis2019} explore the projected radial distribution of X-ray sources in their cluster sample and find evidence for a systematic excess of counts at a projected radius of $\approx2.5\,R_{500}$. In this paper, we build a semi-empirical model of the radial distribution of AGN in massive dark matter halos identified in cosmological N-body simulations. Then we compare the predictions of the model with the observational results of \cite{koulouridis2019}, using the principles of forward modelling. In this approach, observational effects such as X-ray flux limits and the {\it Chandra} field of view are included in the modelling to generate simulated datasets that mimic real observations. In that respect, an important part of the simulations is the X-ray selection function of the observations, which measures the probability of detecting X-ray point sources of a given flux as a function of position within the {\it Chandra} footprint. For that reason we choose to re-analyse the Chandra X-ray observations used by \cite{koulouridis2019} with the reduction pipeline presented by \cite{Laird2009} and \cite{Nandra2015}. The key feature of this pipeline is the sensitivity maps that are constructed following methods presented in \cite{georgakakis2008} and quantify to a high level of accuracy the selection function of the detected X-ray sources. 
 
In brief, the reduction uses standard CIAO tasks to analyse the raw {\it Chandra}/ACIS-I imaging data and produce level-2 event files for individual pointings. Multiple observations of the same field are merged to generate a single event file as well as co-added images and exposure maps in four energy bands 0.5--7.0 keV (full), 0.5--2.0 keV (soft), 2.0--7.0 keV (hard) and 4.0--7.0 keV (ultrahard). Sources are detected independently in each of these spectral intervals following a two-pass process. A seed catalogue of candidate sources is first constructed using the CIAO wavelet-based source detection task {\sc wavdetect} at a low detection threshold of $10^{-4}$. Photons (source and background) at the position of each candidate source are then extracted within apertures that correspond to the 70 per cent encircled energy fraction (EEF) radius of the {\it Chandra} point spread function (PSF) at the source position. The expected background level in each aperture and spectral band is also measured after removing the contribution of nearby source photons. Finally, we estimate for each source the Poisson probability that the observed number of photons within the aperture is the result of background fluctuations. The final catalogue in a given spectral band contains those X-ray sources with Poisson probability as defined above $<4 \times 10^{-6}$. X-ray fluxes are determined assuming a power-law X-ray spectrum with photon index $\Gamma= 1.4$ absorbed by the Galactic hydrogen column density appropriate for each field. The pipeline also produces sensitivity maps \citep[see][]{georgakakis2008}, which measure the probability of detecting an X-ray source with a given count rate or flux as a function of position within the surveyed area. In this work we use the one-dimensional representation of the sensitivity map, the X-ray area curve, which provides an estimate of the total survey area in which a source with a given count rate or flux can be detected. 

\begin{figure}
    \centering 
    \includegraphics[trim={0 1.5cm 0 0}, width=.5\textwidth]{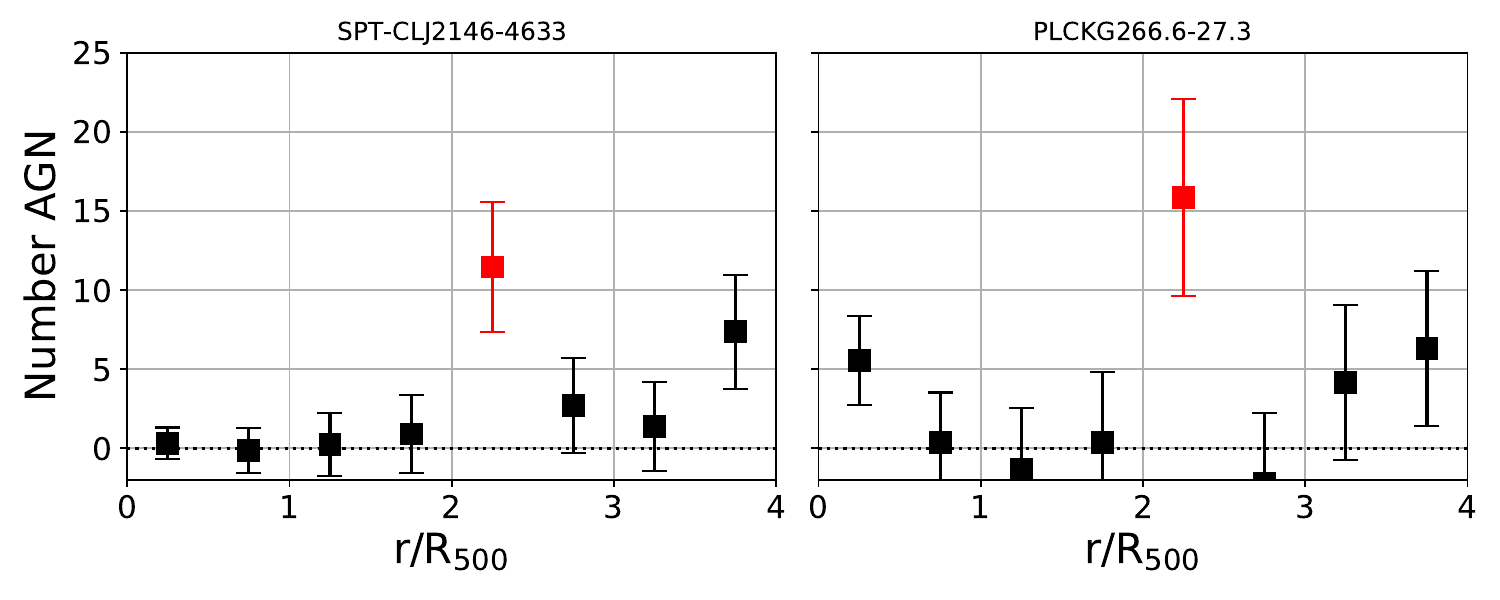}
    \caption{X-ray AGN number counts in the full band (0.5-7.0 keV) for the two selected clusters as function of radial distance in units of $R_{500}$. The expected number of sources in the field have been statistically subtracted from each annulus.}
    \label{fig:rad_distrib_obs}
\end{figure}

Next, we describe the construction of the radial distribution of X-ray sources in each of the clusters in the sample of \cite{koulouridis2019}. We use X-ray sources selected in the full-band (0.5-7\,keV) and group them in radial bins of width $0.5\cdot R_{500}$. These radii are estimated by \cite{Bartalucci2018} by modelling the extended X-ray emission profile of each cluster. The determination of the projected radial distribution of X-ray sources requires the statistical subtraction of the expected number density of foreground/background X-ray sources. This is determined using the number counts as a function of the flux of the extragalactic field X-ray source population, i.e. their $\log N - \log S$ distribution. For a given cluster and $R_{500}$ radial bin, $i$, we first determine the full-band sensitivity curve of the ring with inner and outer radius of $i/2 \cdot R_{500}$ and $(i+1)/2 \cdot R_{500}$ following the methods described in  \cite{georgakakis2008}. We then convolve this with the differential full-band number counts presented by  \cite{georgakakis2008}. This calculation yields the number of extragalactic field X-ray sources (i.e. not associated with the cluster) expected to be detected in the ring under consideration at the depth of the specific {\it Chandra} observation. This expectation value is then subtracted from the observed number of X-ray sources in the ring. The resulting distributions for the clusters PLCKG266.6-27.3 and SPTCLJ2146-4633 are shown in Figure~\ref{fig:rad_distrib_obs}. The two selected clusters are the ones that show the highest overdensity at $2.5\,R_{500}$ in the sample of \cite{koulouridis2019}. This figure also shows that our re-analysis confirms the results of \cite{koulouridis2019}

In the next sections we will use the full-band sensitivity maps of the clusters PLCKG266.6-27.3 and SPTCLJ2146-4633 to forward model the X-ray selection function of real {\it Chandra} observations. The first represents a deep, $\approx 200$\,ks, X-ray observation, the second corresponds to a shallower {\it Chandra} dataset, $\approx 70$\,ks. We will use the sensitivity maps of these observations to explore the impact of different X-ray depths on our results and conclusions. We reiterate that we choose these two clusters because they are the one in the sample of \cite{koulouridis2019} that show the largest amplitude excess counts in their projected radial distributions, which are interpreted as evidence for AGN triggering in their outskirts.

\section{Methodology}\label{sec:methods}
\subsection{The semi-empirical model of AGN and galaxies}\label{sec:SEM}
\begin{figure*}
    \centering 
    \includegraphics[width=1\textwidth]{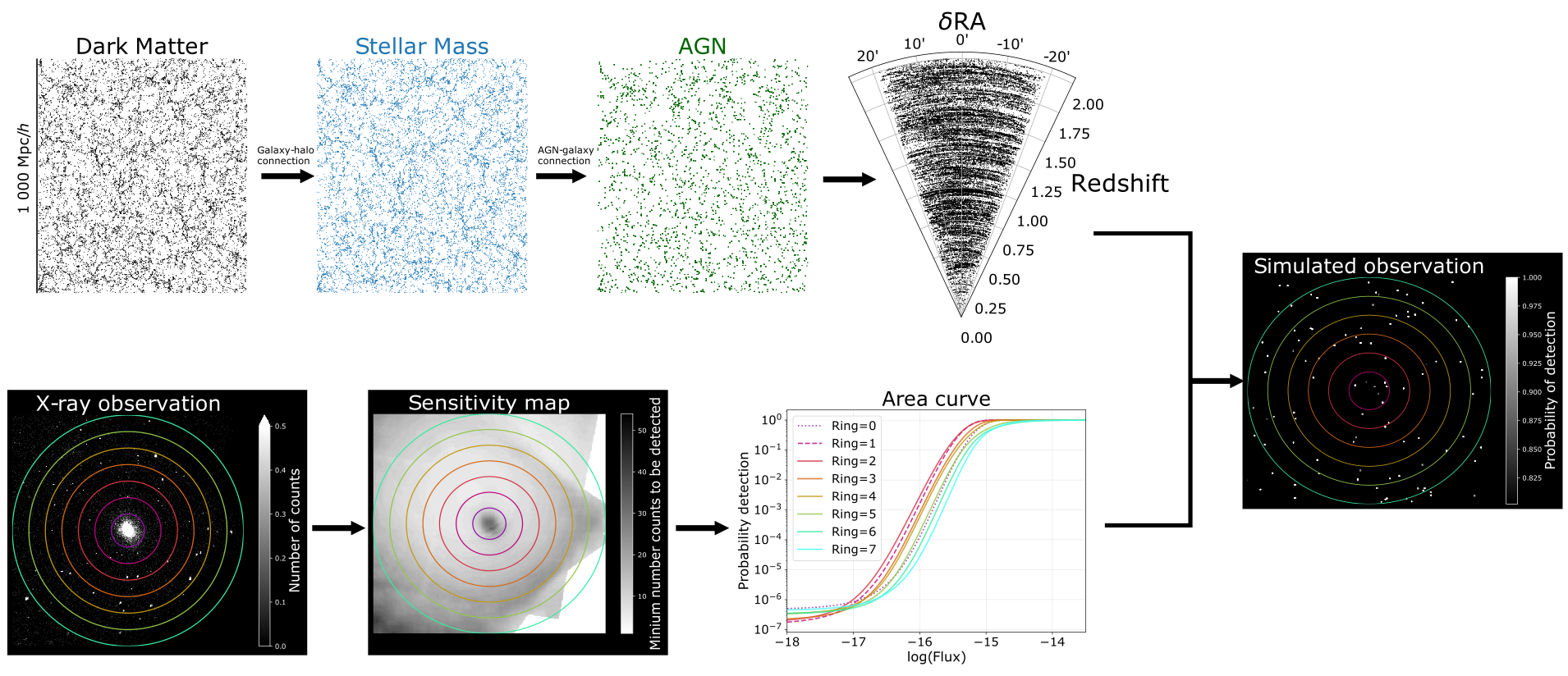}
    \caption{Graphical demonstration of the semi-empirical model that produces realistic simulated observations of AGN in massive halos. The upper branch shows the construction of the AGN mocks. The starting point (left panel) are dark matter only cosmological simulations. Dark matter halos are populated with galaxies using abundance matching techniques (panel labelled stellar mass). Supermassive black-hole accretion events are painted of top of these galaxies using empirical relations that describe the probability of a galaxy hosting an AGN (panel labelled AGN). See Section~\ref{sec:SEM} for details on the mock catalogue construction. The right panel on the upper branch represents a light cone with a field of view 20\,arcmin pointing to a massive halo (i.e. cluster of galaxies) in the mock catalogue as described in Section~\ref{sec:light-cones}. The lower branch shows the derivation of the X-ray selection effects. The starting point are X-ray observations (left panel) of cluster of galaxies presented in Section~\ref{sec:obs}. These are processed to obtain the X-ray sensitivity maps (middle panel) explained in Section~\ref{sec:obs}. The anulii in these panels have radii that are multiples of the quantity $0.5\cdot R_{500}$. These are the anulii used in the observations (see Figure~\ref{fig:rad_distrib_obs}) to study the radial distribution of AGN in clusters. The different colours represent different rings. The right panel shows the area curve for the different rings which describe the probability of detecting a source at different radial distances from the centre of the cluster. The colour of each line corresponds to the anulii shown in the sensitivity map panel. Both branches merge in the panel labelled "Simulated observation", which represents a simulated field of view that includes the selection effects derived from observations.
    }
    \label{fig:sktch_wrkflw}
\end{figure*}
In this section, we describe the development of the semi-empirical model that is used to interpret the observations presented in Section~\ref{sec:obs} on the radial distribution of AGN in massive clusters of galaxies. The semi-empirical approach is a flexible data-driven method that produces realistic mock catalogues of galaxies \cite[e.g.][]{Moster2018, behroozi19_um, Grylls2019} and/or AGN \cite[e.g.][]{Comparat2019, Comparat2020, Seppi2022, Zhang2022}. By construction, such mocks obey observed properties of galaxy and/or AGN populations, e.g. the stellar mass function, the star formation main sequence at different cosmic times and the AGN luminosity function. In contrast with other modelling methods, such as hydrodynamical simulations or semi-analytical models, the semi-empirical approach does not rely on a set of recipes to describe the physical mechanisms that regulate galaxy/AGN evolution. Instead, empirical assumptions are made, e.g. the stellar mass of a galaxy correlates with halo mass, which can usually be described by few parameters. Because of its simplicity, the semi-empirical approach is ideal for testing specific hypotheses by comparing simulations with observations. It is this latter point that motivates the use of the semi-empirical approach in our analysis, instead of more complex and physically driven simulations of massive clusters of galaxies \citep[e.g.][]{Cui2018}.

In this work we follow the methodology described in \cite{munozRodriguez2023} to construct AGN mock catalogues. In brief, the backbone of the model is a dark matter only N-body simulation. It provides the dark matter halo structure within which galaxies form and evolve. We choose to use the MultiDark PLanck2 \citep[MDPL2, ][]{MultiDark2016} because it is one of the largest volume, high resolution and public cosmological simulations. It has a box size of 1000 Mpc/$h$, a mass resolution of $1.5\times10^9\,$M$_\odot/h$ and 3\,840$^3$ ($\sim$57$\cdot$10$^9$) particles. Dark matter halos are populated with galaxies using abundance matching techniques. In particular, we use the \textsc{UniverseMachine} model of \cite{behroozi19_um} implemented for the MDPL2 dark matter N-body simulation. This model assigns galaxies to halos by parameterising the star-formation rate (SFR) in terms of halo properties (mass, accretion history and cosmic time). By integrating the SFR across the halo history it is possible to predict observables that are compared with real observations, including for example, the stellar mass function and the evolution of the cosmic star-formation rate density. The best model is found by iterating the comparison between predictions and observations to explore the model parameter space. The end product of \textsc{UniverseMachine} are catalogues of dark matter halos, each of which is assigned a galaxy stellar mass and a SFR. 

Following \cite{munozRodriguez2023}, an AGN luminosity is assigned to each galaxy in \textsc{UniverseMachine} using observational measurements of the AGN Specific Accretion Rate Distribution \citep[SARD;][]{bongiorno12_cosmos, aird12_primus,Bongiorno2016,age17_sar,aird18_sar}. This quantity describes the probability of a galaxy hosting an accretion event onto its supermassive black hole with specific accretion rate (SAR) $\lambda_{SAR} = L_X/M_{\star}$, where $L_X$ is the AGN luminosity (in this case at X-rays) and $M_\star$ is the stellar mass of the host galaxy. The observationally-derived SARDs are used to assign accretion events to mock galaxies in a probabilistic way \citep[e.g.][]{age19_mock, aird21_mock,munozRodriguez2023} and therefore include mock AGN in the \textsc{UniverseMachine} boxes. The fundamental assumption of this step is the lack of a physical connection between the accretion events and the environment, i.e., the AGN incidence is stochastic in nature and independent of the halo mass. The process of constructing the galaxy and AGN semi-empirical model described above is illustrated in the first three panels from left to right on the upper branch of Figure~\ref{fig:sktch_wrkflw}.

The catalogues of mock AGN and galaxies produced above need to be further processed to mimic observations of the real Universe and allow the comparison with observational results in a forward modelling manner. The essential step for achieving this is the projection of the boxes onto the sky plane to construct light cones as in \cite{munozRodriguez2023}. However, the light cone requirements of the present work are very different from those in \cite{munozRodriguez2023}. As a result the construction of this product deviates from our previous study and is described in detail in the next section.

\subsection{Light-cones}\label{sec:light-cones} 
\begin{figure}
    \centering        
    \includegraphics[width=.49\textwidth]{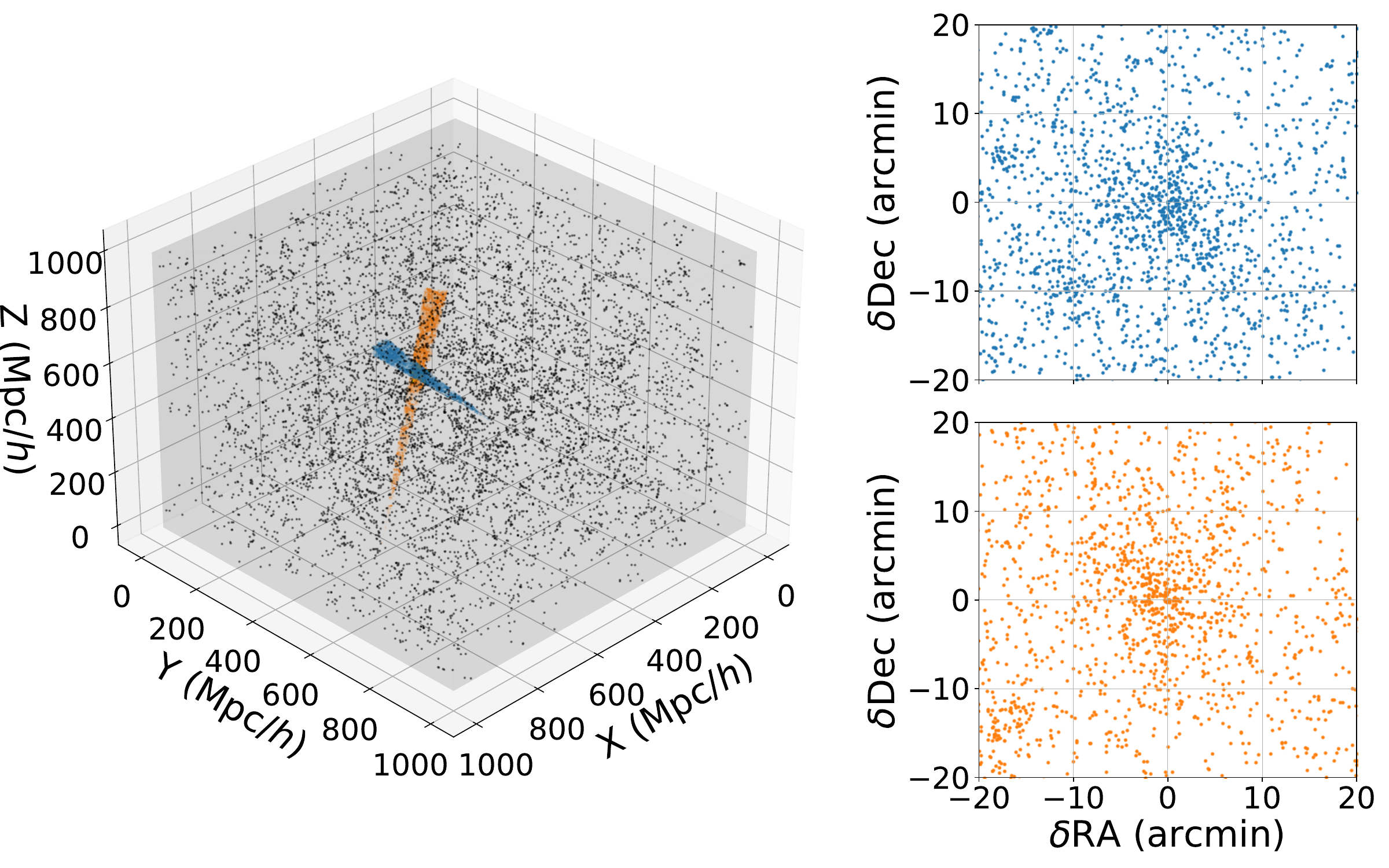}
    \caption{Examples of two light cones that intersect different sight-lines through the cosmic web. The left panel shows a 3-dimensional projection of the space. The shaded region represents the boundaries of a MDPL2 box. Black points are dark matter haloes with masses $>10^{14.25}$~M$_\odot/h$ in the simulation. This limit is chosen to help visualisation. Blue and orange points represent the two different light-cones with all the haloes (irrespective of their mass) within their solid angle. The right panels show the corresponding projections of each of the light cones. Each point on the right set of panels represents a dark matter halo.}
    \label{fig:diff_project}
\end{figure}

In this work, we explore the projected radial distribution of X-ray-selected AGN in galaxy clusters and how this is affected by sample variance. At the simulation level, this is investigated by generating light cones of massive dark matter halos whose sight lines probe different paths through the cosmic web. This is demonstrated in Figure~\ref{fig:diff_project} which shows two different sight lines to a particular halo (left panel). The corresponding projected structures along these sight lines are also shown in the figure.  In the next sections, we first discuss the general approach for constructing light cones (Section~\ref{sec:Deep-beam_light-cone}) and explain how this is modified to allow more freedom in the choice of sight lines to a particular halo (Section~\ref{sec:cut_and_paste}). 

\subsubsection{General light cone construction}\label{sec:Deep-beam_light-cone}
Extragalactic surveys are typically characterised by a finite field of view and a flux limit at some waveband that allows the detection of astrophysical sources (galaxies or AGN) over a wide range of redshifts. Dark matter N-body simulations like MDPL2 have a  finite box size, which when projected onto the sky plane samples only a limited redshift range\footnote{For example, the centre of a box with a length-size of 1~Gpc/$h$ at $z=1$ corresponds to a comoving distance of $D_{c,centre}\sim2300$~Mpc. The bottom and top faces lie at $D_{c, bottom}\sim 1800$~Mpc and $D_{c, top}\sim 2800$~Mpc respectively. These distances correspond to the redshift range $z\sim0.72-1.3$.}. Producing mocks over a wide redshift interval requires that simulated boxes are used as building blocks to construct a 3-dimensional (3D) pavement. The stack of boxes can be extended from $z=0$ to an arbitrary maximum redshift ($z_{max}$) by selecting an appropriate number of boxes. A wide range of redshift, however, corresponds to a significant look-back time, during which the structure of the Universe evolves strongly. This effect can be captured by selecting different dark matter simulation boxes that correspond to different redshifts. They represent snapshots of the cosmic web at distinct times during the lifetime of the Universe. Using different snapshots and stacking them we construct catalogues that describe the evolution of the structure in the Universe over a wide range of redshifts. The skeleton of these catalogues can be described as an onion-shell structure where each slice corresponds to a different snapshot.  A potential issue with this approach is that since distinct snapshots represent the same volume of the simulated Universe, the positions of specific structures are correlated between different boxes. This is known as the repetition problem. Diverse alternatives have been proposed in the literature to address this limitation. We implement random tiling, which decorrelates relative positions between different boxes by rotating them along the main axes of the box when they are stacked \citep[see][]{momaf_blaizot2005,Bernyk2016}. 

\begin{figure*}
    \centering 
    \includegraphics[width=1\textwidth]{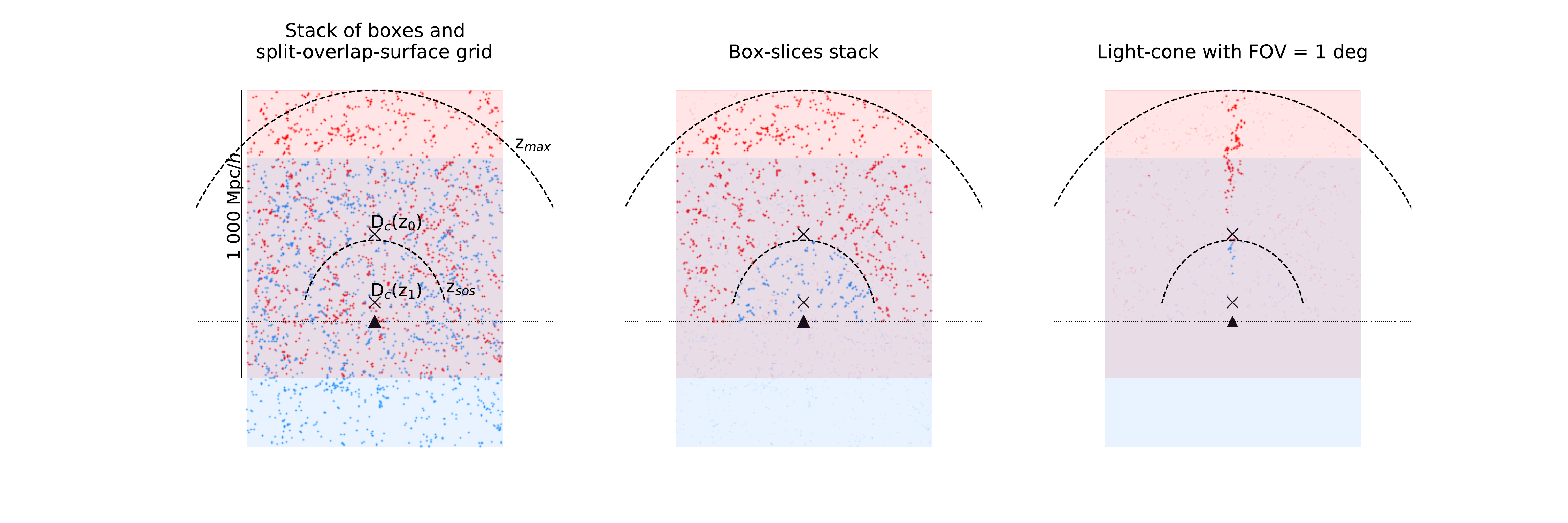}
    \caption{Sketch that illustrates the step-by-step construction of a light-cone for an observer located underneath the centre of the N-body simulation box. The observer's positions is indicated with a black triangle that lies on the plane indicated with the solid horizontal line. In all panels we show a stack of 2 boxes at different snapshot redshifts, coloured differently for illustration purposes. The red shading is for the higher redshift box (relative to the observer) whereas the blue colour correspond to the lower redshift box. The extent of the shaded regions indicates the size of the boxes. The crosses indicate the centres of each box, which in this example are also used as pivot points (see Section~\ref{sec:light-cones} for details). 
    The dots within each box correspond to dark matter halos with masses $\rm{M}_{halo}>10^{13.5}~\rm{M}_\odot/h$ and have the same colour (blue or red) as that of the box they belong to. The black dashed curves represent the iso-redshift surfaces relative to the observer. These define the split-overlap-surfaces used to select halos from the different boxes (see Section~\ref{sec:light-cones} for details). The construction of the light-cone proceeds from left to right: (i) First, we offset each box in the vertical axis so that the redshift of its pivot point (cross) relative to the observer equals the snapshot redshift of the box; (ii) a set of split-overlap-surface is defined with respect to the grid of boxes (dashed curves); (iii) the set of split-overlap-surfaces is used to remove dublicate halos in the overlap region of the two boxes (middle panel): below the lower dashed curve only blue halos are kept, whereas between the lower and upper dashed curves only red halos remain; (iv) the field-of-view is applied to the box-slices (right panel), keeping only halos that are within a user defined solid angle.
    }
    \label{fig:sktch_wrkflw_lcs}
\end{figure*}

The origin of the reference system of a given simulation box is assumed to be located at the centre of the box. The cartesian coordinates of the individual objects therefore, take values in the range [-L$_{box}$/2, L$_{box}$/2], where L$_{box}$ is the length-side of the simulated volume. The hypothetical observer is located onto the XY plane at $Z=0$. Its precise position on the plane can be almost arbitrary. Constraints are discussed in Section~\ref{sec:impl_ths_wrk}. The box is located at a comoving distance that corresponds to a reference redshift ($z_{ref}$) with respect to the position of the observer. This is achieved by offsetting the box along the Z-axis by $\Delta z$ defined as

\begin{equation}
    D_c(z=z_{ref}) = \sqrt{\rm{x}^2 + \rm{y}^2 + (\rm{z + \Delta z})^2},
    \label{eq:com-dist_redshift}
\end{equation}

\noindent where $D_c(z=z_{ref})$ is the comoving distance that corresponds to $z_{ref}$. The coordinates in the equation above are defined as x = x$_{pp}$ - x$_{RS}$,  y = y$_{pp}$ - y$_{RS}$ and z = z$_{pp}$, where $RS$ are the coordinates of the observer and $PP$ are the coordinates of the pivot point. The latter are defined as the coordinates of a point within the box that has a distance of exactly $D_c(z=z_{ref})$ from the observer. The exact location of the pivot point within the box is a free parameter although it is typically chosen to be the centre of the box. The $z_{ref}$ usually corresponds to the reference redshift of the snapshot. Deviations from these norms are discussed in Section~\ref{sec:impl_ths_wrk}.

The stacking of boxes requires some overlap between consecutive boxes to avoid empty volumes which would generate an incomplete light cone. This is achieved by imposing the condition $D_c(z_i) - D_c(z_{i+1}) < L_{box}$, i.e., the comoving distance between the centres of consecutive boxes should be smaller than their comoving length. However, the overlap produces artificial overdense regions because the same volume contains objects from two different boxes. This is demonstrated on the left panel of Figure~\ref{fig:sktch_wrkflw_lcs}, which shows the stack of two boxes. The intersecting volume contains the individual halos of each box and therefore, it has an artificially enhanced density. We address this issue by defining a boundary surface of constant comoving distance (or redshift) relative to the observer. We refer to this as the split-overlap-surface ($sos$). It determines which objects are adopted from each box. Above the surface, only halos from the box on the top are kept. Whereas below the surface, only objects from the bottom box are retained. This is illustrated in the middle panel of Figure~\ref{fig:sktch_wrkflw_lcs}, where the lower curved line represents the split-overlap-surface.

The split-overlap-surfaces define a set of box slices, i.e. the onion shell structure of the light cone. The stacking of the simulation boxes to construct the light cone follows a top-to-bottom approach: the pivot point of the highest redshift box is defined and the appropriate offset relative to the observer is applied to it. The sight line between that pivot point and the observer define the axis of symmetry of the light cone. Lower redshift boxes are then added underneath the first one by defining appropriate pivot points and split-overlap-surfaces. The relative angle between the objects in the box slices and the selected sight line is calculated. This angle can be decomposed into a right-ascension and declination on the unit sphere. The redshifts associated with the individual halos correspond to their comoving distances with respect to the observer. Then the input field of view of the light-cone is applied by rejecting objects with angular distances larger than the adopted solid angle. This is illustrated on the right panel of Figure~\ref{fig:sktch_wrkflw_lcs}, where only objects within the limits of the field of view are included.  

\subsubsection{Cut-and-paste method}\label{sec:cut_and_paste}
\begin{figure*}
    \centering 
    \includegraphics[width=.8\textwidth]{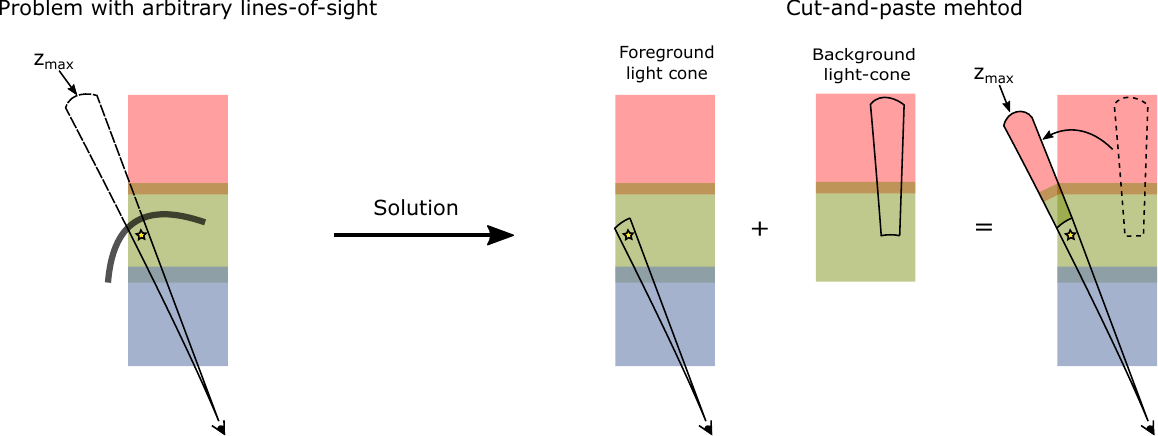}
    \caption{A sketch that illustrates the problem of using arbitrary lines of sight when constructing light cones that are forced to intercept a particular halo. The panels show a stack of 3 boxes at different snapshot redshifts. Each of the boxes is  coloured differently (blue,  green or red) for illustration purposes. The blue and red shadings correspond to the lowest and highest redshifts of the stack. The position of the observer is shown at the bottom of each stack of boxes with the eyeball graphic. The left panel shows an example of a tilted sight line that is forced to contain the position of a halo marked with the star symbol in the green box. This sight line hits the boundaries of the stack of boxes before reaching the expected maximum redshift $z_{max}$ (see Section~\ref{sec:Deep-beam_light-cone}). Such a light cone is clearly incomplete. The right panel visualises a solution to this issue that  is based on the construction of two independent light cones. The first one encompasses the region between the observer and the last redshift slice where the light cone is complete. We refer to this component as the foreground light cone. A second independent light cone is then constructed that extends from the previous complete redshift to $z_{max}$. We refer to  this component as background light cone. Finally, both light cones are aligned by matching the lines of sight. This is indicated in the far right panel. The arrow shows the rotation that needs to be applied to the background light cone to align it to the foreground one.}
    \label{fig:sktch_cut_paste}
\end{figure*}

For our specific application it is necessary to construct light cones that intersect a particular halo position (i.e. that of a massive cluster) at a comoving distance from the fiducial observer that corresponds to a fixed redshift. Therefore the pivot point of the box that contains this particular halo is set to the Cartesian position of this halo. In this case, the methodology described in the previous section has a limitation that is demonstrated in the left panel of Figure~\ref{fig:sktch_cut_paste}. The sight line to the target object may intersect the boundaries of a box in the stack before reaching the maximum redshift of the light cone. We address this issue by modifying the methodology described in Section~\ref{sec:Deep-beam_light-cone}.

The solution is based on the construction of two independent light cones, as illustrated on the right panel of Figure~\ref{fig:sktch_cut_paste}. The first light cone extends from the observer at $z=0$ up to the redshift surface where the line of sight intersects the boundaries of the stack. This is referred to as the foreground light cone. The second light cone expands from the last redshift surface of the foreground light cone up to the maximum redshift, $z_{max}$, and has a different orientation compared to the first one to ensure that no box boundaries are hit out to $z_{max}$. This is referred to as the background light cone. The line of sight of each light cone is specified by the tuple of positions defined by the target object and the observer. For the foreground light cone the target object is a specific selected halo in the simulation and the observer position is randomly generated on the XY plane. In the case of the background light cone, the observer is located underneath a randomly selected position of the last box of the stack. Each of the light cones are then assembled following the approach described in Section~\ref{sec:Deep-beam_light-cone}. Clearly the axes of symmetry are misaligned since they are built independently and point to different directions. Nevertheless, for the light cone construction, the only relevant quantity is the relative angles of an object with respect to the axis of the light cone, i.e. $\delta RA$ and $\delta Dec$. These are independent of the direction of the light cone axis. Hence, they can be used to align the two independent light cones, the foreground and background ones, to point to the same direction. We refer to this methodology as cut-and-paste. 

\subsection{Implementation for this work: Simulating a realistic set of observations}\label{sec:impl_ths_wrk}
We use the implementation of \textsc{UniverseMachine} \citep[][]{behroozi19_um} on the MDPL2 \citep[][]{MultiDark2016} cosmological simulation with a side of 1\,Gpc/$h$. We select a total of 12 \textsc{UniverseMachine} boxes at different snapshot redshifts chosen to cover the redshift range $z=0-3$ in steps of $\sim1$ Gyr. Mock AGN are assigned to \textsc{UniverseMachine} galaxies using the SARDs of either \cite{age17_sar} or \cite{aird18_sar}. Our baseline simulations use as reference the observations of the cluster PLCKG266.6-27.3 with a mass of $M_{500}^{SZ}=8.5\times10^{14}$\,M$_\odot$ at a redshift of $z = 0.97$ \citep[see][]{Bartalucci2018}. This is because PLCKG266.6-27.3 is the cluster in the sample of \cite{koulouridis2019} that shows the highest excess of X-ray sources at a projected radial distance of 2.5~$R_{500}$. The mock \textit{Chandra}/ACIS-I observations of PLCKG266.6-27.3 use massive halos drawn from the \textsc{UniverseMachine} box at a snapshot redshift of $z=0.94$, i.e., similar to the redshift of the real cluster. There are 10 halos in that box with M$_{500c}$>$5\times 10^{14}$\,M$_\odot$  \footnote{\textsc{UniverseMachine} provides only virial halos masses. We convert these values to $500$ critical, using a mean halo concentration $c=0.7$ \citep{Ludlow2014Mass_concent_clust}.} , i.e. similar to the limiting mass of the \cite{koulouridis2019} sample. 
The light cones are constructed to target the most massive halo in the simulation box with a mass of M$_{500c}\sim7.5\times 10^{14}$\,M$_\odot$ (\textsc{UniverseMachine} identification number id=7830644447). We study the impact of halo mass on our results and conclusions by  also constructing light cones that pass through a second less massive halo (\textsc{UniverseMachine} identification number id=7793510527) with mass M$_{500c}\sim5\times 10^{14}$\,M$_\odot$. In the next section we show that our analysis is not sensitive to the choice of the massive halo used to simulate light cones of clusters of galaxies.

We generate 100 lines of sights pointing to each of these two clusters with a field of view  set to 20~arcmin diameter, which mimics the \textit{Chandra}/ACIS-I observations. For each simulated observer we produce the projected radial distribution of mock X-ray selected AGN by splitting the field of view in 8 concentric rings. The $i$-th ring is assigned an outer radius $r_i=i\cdot 0.5~R_{500}$ from the cluster centre. Mock galaxies, and therefore, AGN of the light cone are assigned to a ring depending on their projected radial distance relative to the cluster centre. The application of observational selection effects onto the simulation requires the estimation of AGN fluxes. They are assigned to the X-ray AGN luminosities by assuming a power-law spectral shape with photon index $\Gamma=1.4$. Applying the corresponding sensitivity curve of each ring (see Section~\ref{sec:obs}) a probability of detection is assigned to each source based on its flux. The total number of AGN per ring is calculated as the sum of probabilities of the AGN within the ring. The final product of this process are 100 AGN radial distributions that represent the 100 simulated lines of sight. The background is statistically subtracted as in observations (see Section~\ref{sec:obs}). We calculate the expected number of background/foreground AGN within each ring by simulating field (i.e. off cluster) observations. We generate a field sample by constructing 100 light cones that point to 100 random locations in the box at $z=0.97$, i.e. the same box as the simulated clusters. For simplicity we always locate the observer underneath the random target point. The same set of split overlap surfaces used for the clusters is also applied to the field observations. This is because we require the same redshift structure in both samples. We calculate the AGN distribution for this sample following the same steps described for the clusters. Each AGN is assigned to one ring using its projected radial distance with respect to the centre of the field. Detection probabilities are assigned to the AGN using the corresponding area curve. Finally, the expected field value is calculated as the average number of AGN per radial bin in the 100 simulations. 

\section{Results}\label{sec:results}
\subsection{AGN Halo occupation predictions}   
\begin{figure*}
    \centering 
     \includegraphics[width=1\textwidth]{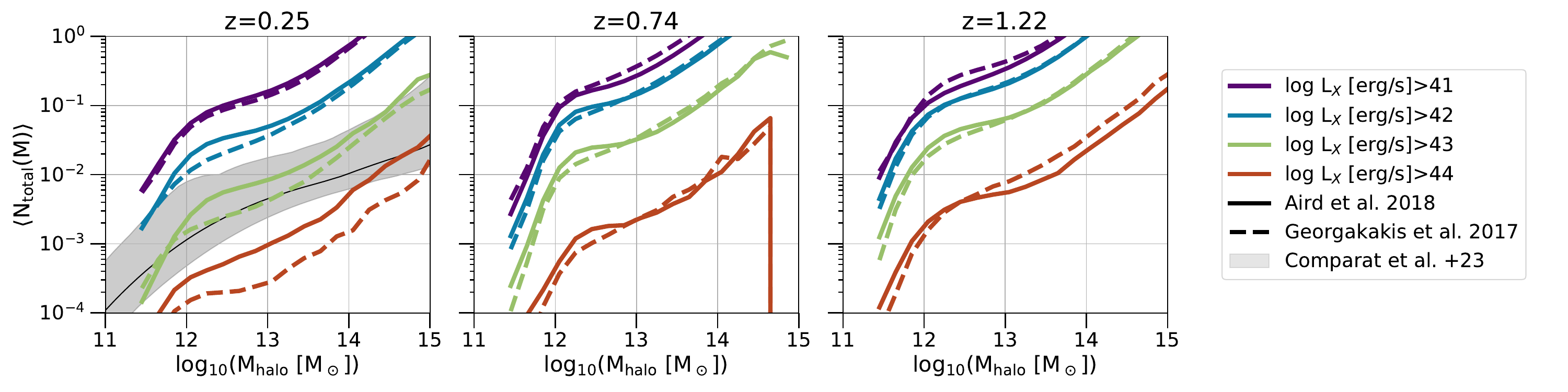}
    \caption{Semi-empirical model predictions for the AGN HOD at different X-ray luminosity thresholds (indicated in the legend) and redshifts (indicated in the title of each panel). The different models of the specific accretion rate distribution used in our semi-empirical approach are indicated with different line styles. Solid lines correspond to \citet{aird18_sar} whereas dashed lines are for \citet{age17_sar} (see Section~\ref{sec:SEM} for details). At the redshift panel $z=0.25$ we also compare the predictions of the model with the observational results of \citet[][]{2023Comparat}. The black solid line correspond to the best-fit AGN HOD from this study. The 1$\sigma$ uncertainties are shown with by the grey shaded region.}
    \label{fig:HOD_pred}
\end{figure*}
  
Before focusing on the radial distribution of AGN in massive clusters of galaxies at $z\approx1$, we present general predictions of our semi-empirical model (see Section \ref{sec:SEM}) on the halo occupation distribution (HOD) of X-ray selected AGN. Such model predictions can be compared against current and future observational constraints to gain insights into the triggering mechanisms of accretion events onto SMBHs at different environments. We reiterate that our semi-empirical model is build upon the fundamental assumption that the clustering of AGN follows that of their host galaxies. The latter is included in the modelling of the galaxy-halo connection as implemented by  \textsc{UniverseMacine}. Any discrepancies between observed AGN HODs and the semi-empirical model predictions would question the assumption above, thereby pointing to environmental effects that modulate the incidence of AGN in halos \citep[e.g.,][]{munozRodriguez2023}. We also remind the reader that the AGN-galaxy connection approach presented in Section \ref{sec:SEM} produces mock AGN catalogues that are consistent with the observed 2-point correlation function of different AGN samples that span a range of accretion luminosities and redshifts \citep{age19_mock}. In that respect our semi-empirical model is consistent with the large scale distribution of AGN in the Universe. 

The AGN HOD, $\langle N(L_X)| M \rangle$, is defined as the mean number of AGN brighter than the luminosity $L_X$ in dark matter halos of given mass, $M$. Because of the different halo types (central or satellites) the HOD is usually expressed as a sum of two terms

\begin{equation}
    \begin{split}
    \langle N(L_X)| M \rangle \;&= \langle N_{cen}(L_X)| M \rangle + \langle N_{sat}(L_X)| M \rangle ,\\ 
     \langle N_{cen}(L_X)| M \rangle &= f_{A}\cdot\frac{N_{AGN,cen} (M,\;L_X)}{N_{cen} (M)},\\
     \langle N_{sat}(L_X)| M \rangle &= \frac{N_{AGN,sat} (M_{par}=M, \; L_X)}{N_{cen} (M)},
  \end{split}
  \label{eq:HOD}
  \end{equation}

\noindent where $\langle N_{cen}(L_X)| M \rangle$, $\langle N_{sat}(L_X)| M \rangle$ is the mean number of AGN brighter than $L_X$ in parent halos of mass $M$ that are associated with central and satellite galaxies respectively. $f_{A}$ is a normalisation factor that represents the fraction of active galaxies with respect to the full population.

Figure~\ref{fig:HOD_pred} shows our HOD predictions for different X-ray luminosity cuts and redshifts. These are estimated by populating the \textsc{UniverseMachine} boxes at the corresponding redshifts with AGN and then applying Equation~\ref{eq:HOD}. At fixed luminosity threshold the HOD normalisation increases towards higher redshift. This is the result of the strong increase of the AGN space density to redshift $z\approx2-3$. 
Also, the HOD normalisation decreases toward higher luminosities as a result of the form of the AGN X-ray luminosity function. Figure~\ref{fig:HOD_pred} further shows that both specific accretion rate models used to seed galaxies with AGN produce similar HOD results. However, the differences between the two models are stronger at the lowest redshift bin ($z\sim0.25$) and toward higher X-ray luminosities. These discrepancies are related to the modelling of the observed specific accretion rate distributions by \cite{age17_sar} and \cite{aird18_sar} as already discussed in \cite{munozRodriguez2023}. 

Figure~\ref{fig:HOD_pred} also compares our semi-empirical model predictions with recent results on the HOD of X-ray selected AGN in the eROSITA eFEDS field \citep{2023Comparat}. This sample selects AGN in the redshift interval $z=0.05-0.55$ (average of 0.34) and mean X-ray luminosity in the 0.5--2\,keV band of $\approx 10^{43}\rm erg\, s^{-1}$. Two clustering statistics, the 2-point correlation function and weak lensing, are applied to this sample to measure the AGN halo occupation
distribution. We caution that the normalisation of the AGN HOD, i.e., parameter $f_A$ in Equation~\ref{eq:HOD}, cannot be inferred from the observations \citep[][]{allevato21,2022Carraro}. Instead this important quantity is determined post-processing based on knowledge of the AGN X-ray luminosity function and halo mass function at the redshifts of interest \citep[e.g.][]{2023Krumpe}. For the comparison we fixed $f_A=0.01$, which correspond to the duty cycle of central galaxies derived from the specific accretion rate distributions at similar redshift and luminosity threshold to those used by \cite{2023Comparat}. Although the uncertainties of the observations are large, the best-fit AGN HOD increases with increasing halo mass slower (i.e. flatter slope) than the model predictions. This suggests a suppression of AGN activity toward massive halos, i.e., cluster of galaxies, at $z\approx0.25$ compared to the semi-empirical model expectation. This is in line with the arguments presented in \cite{munozRodriguez2023} based on the forward modelling of the observed fraction of AGN in massive clusters of galaxies. In that study it is found that the same semi-empirical model presented in Section~\ref{sec:SEM} overpredicts the incidence of AGN among galaxies in low redshift clusters. 

\subsection{Observed overdensity of AGN}
\begin{figure}
    \centering 
    \includegraphics[width=.5\textwidth]{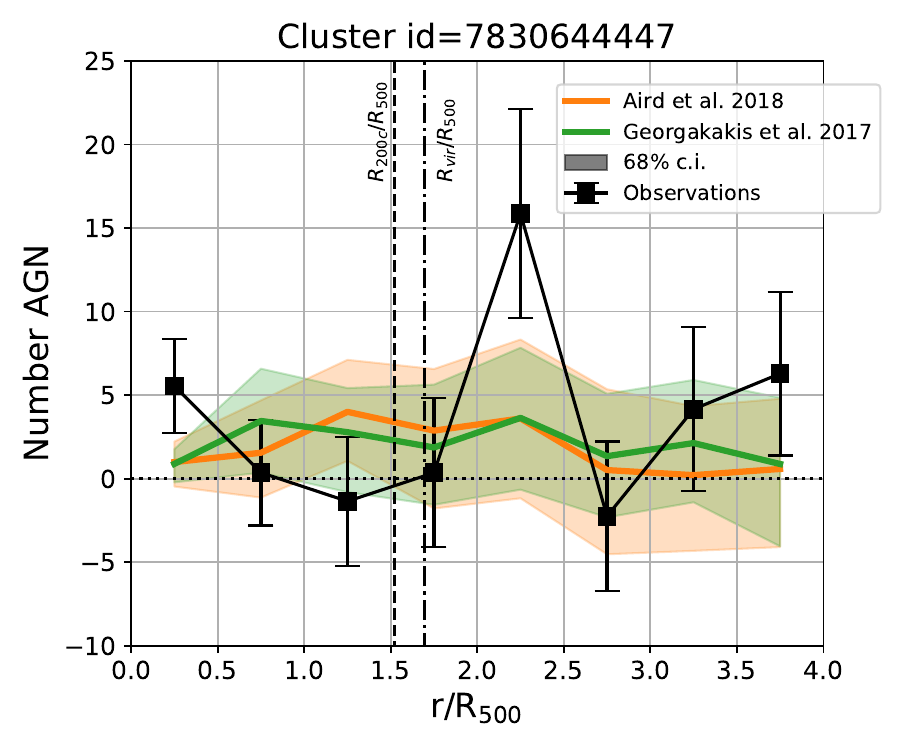}
    \caption{The observed projected radial distribution of X-ray selected AGN of the cluster PLCKG266.6-27.3 (black points connected with the solid black line) is compared with the semi-analytic model predictions (colored lines and shaded regions). The green and orange solid lines show the mean radial distributions of simulated X-ray AGN assuming the \citet{age17_sar} and \citet{aird18_sar} SARDs, respectively. The average at each radial bin is estimated from the 100 light cone realisations described in Section \ref{sec:impl_ths_wrk}, which point to the massive halo ($M_{500c}\approx 7.5\times 10^{14}$~M$_\odot/h$) with id=7830644447 in \textsc{UniverseMachine}. The light green and light orange shaded regions correspond to the 68\% confidence intervals around the mean value at each radial bin. This scatter represents the (cosmic) variance among the 100 light cone realisations. The vertical lines represent the correspondent $R_{200,c}$ (dashed) and $R_{vir}$ (dashed-dotted) normalised to the $R_{500c}$ of the cluster.}
    \label{fig:rad_distrib}
\end{figure}

Next we compare the projected radial distribution of X-ray selected AGN in the cluster PLCKG266.6-27.3 (see Figure~\ref{fig:rad_distrib_obs}) with the predictions of the semi-empirical model described in the previous sections. Figure~\ref{fig:rad_distrib} shows this comparison for two versions of the model based on either the \cite{aird18_sar} or the \cite{age17_sar} specific accretion rate distributions. The MDPL2 halo selected to represent the cluster PLCKG266.6-27.3 has a catalogued identification number id = 7830644447 in \textsc{UniverseMachine} and a halo mass of M$_{500c}\sim7.5\times 10^{14}$\,M$_\odot$. At fixed $R_{500}$ radial bin, Figure~\ref{fig:rad_distrib} plots the mean of the model predictions and the corresponding 68\% confidence intervals. These quantities are determined from the radial distributions of individual fiducial observers. The scatter around the mean (68\% confidence interval) therefore provides an estimate of the sample variance, i.e., the fact that different observers see different structures along their corresponding lines of sight to the cluster.

The simulations predict, on average, a flat radial distribution independent of the adopted specific accretion rate model used to seed galaxies with AGN. This is an expected behaviour of the model, which assumes that the incidence of AGN in galaxies (i.e., the probability of triggering an accretion event onto a SMBH) is independent of environment. As a result, there is no special physical scale in the model at which an overdensity of AGN should be expected. A striking feature in Figure~\ref{fig:rad_distrib} is the large scatter around the mean at fixed $R_{500}$ radial bin. In that respect, it is interesting that within the 1$\sigma$ sample variance uncertainty the predictions of the models are in agreement with the observations. We reiterate that the origin of this scatter is the diversity of projected structures along the line of sight of different observers. It is therefore interesting to explore whether individual simulated observers (i.e., individual light cone realisations) see X-ray AGN radial distributions with features similar to the observed ones, i.e. excess counts. 

\begin{figure*}
    \centering 
     \includegraphics[width=1\textwidth]{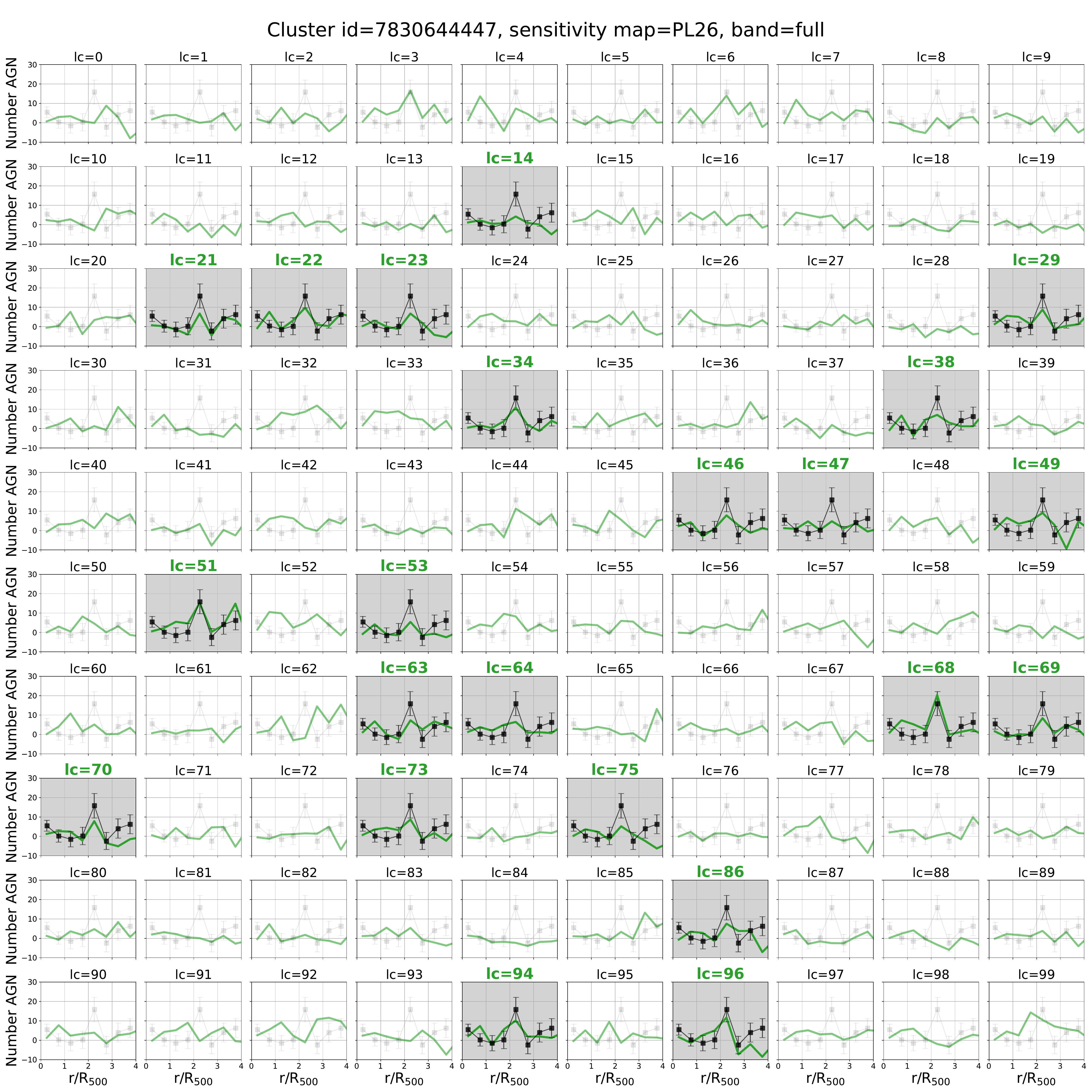}
    \caption{Each panel plots the radial distributions of mock X-ray AGN (green circles connected with green solid lines) for each of the 100 individual light-cone realisations that point to the massive halo ($M_{500c}\approx 7.5\times10^{14}$~M$_\odot/h$) with id=7830644447 in \textsc{UniverseMachine}. The semi-empirical model predictions shown in each panel use the specific accretion rate distribution of \citet{age17_sar} to seed galaxies with AGN. The observed projected radial distribution of X-ray selected AGN of the cluster PLCKG266.6-27.3 is shown with the grey/black squares connected with the solid grey/black lines (see Figure \ref{fig:rad_distrib_obs} and Section \ref{sec:obs}). The light-cone realisations that reproduce the observed peak at the distance of 2.5~$R_{500}$ are highlighted by (i) making the observational data points and connecting lines black, (ii) using bold green characters for the light-cone incremental number at the top of the corresponding panel and (iii) change in the background colour from white to grey.}
    \label{fig:indiv_rad_distrib}
\end{figure*}

Figure~\ref{fig:indiv_rad_distrib} shows the radial distributions for each of the 100 fiducial observers. Eyeballing each of these realisations would identify a few that show an X-ray AGN overdensity at the radial distance of 2.5~$R_{500}$ relative to the neighbouring bins. This approach however, is subjective and therefore we define a set of quantitative criteria to select light cones with excess counts. The adopted conditions that should be simultaneously met are

\begin{equation}
    \begin{split}
        &N_{AGN} (r_i)>0.5\sigma_{r_i}\\
        &N_{AGN} (r_{i\pm1}) < 1 \sigma_{r_{i\pm1}}\\
        &N_{AGN} (r_{i}) >  N_{AGN} (r_{j}), i\neq j
    \end{split}
    \label{eq:overdensity_cond}
\end{equation}

\noindent where $N_{AGN}(r_i)$ is the number of AGN at the ring $i$, $i$ indicates the ring of the overdensity (i.e. $r=2-2.5~R_{500}$), $\sigma_{i}$ is the scatter in the correspondent radial bin, and $i\pm1$ indicate the previous and subsequent ring (i.e., $r=1.5-2~R_{500}$ and $r=2.5-3~R_{500}$ respectively). We reiterate that this criterion is empirically motivated, i.e. it is tuned to broadly select simulated radial distributions similar to the observations. Therefore, visual inspection of Figure 9 may reveal either simulated radial distributions that fulfil the criteria but show marginally significant peaks at r=2--2.5~$R_{500}$ (e.g., lc=14, 38 or 47) or, conversely, realisations that show excess counts at that ring but are not picked by the criteria (e.g., lc=3, 6 or 44). We acknowledge these issues, which on the other hand, emphasise the necessity of having a quantitative, objective and reproducible approach for selecting simulated projected radial distributions. Hence, equation \ref{eq:overdensity_cond} provides a basis for the quantitative assessment of the frequency of AGN overdensities in their projected radial distribution. Figure~\ref{fig:indiv_rad_distrib} highlights the realisations that fulfil the above criteria. It demonstrates that $\sim$20\% of the observers reproduce similar peaks as in the observations. This frequency is only mildly sensitive to the adopted criteria. We therefore conclude that sample variance needs to be taken into account when interpreting the radial distribution of AGN in massive clusters of galaxies. A non-negligible fraction of our simulation realisations can reproduce the most extreme cluster, in terms of excess AGN, of the sample of \cite{koulouridis2019}.

\begin{figure}
    \centering 
    \includegraphics[width=.5\textwidth]{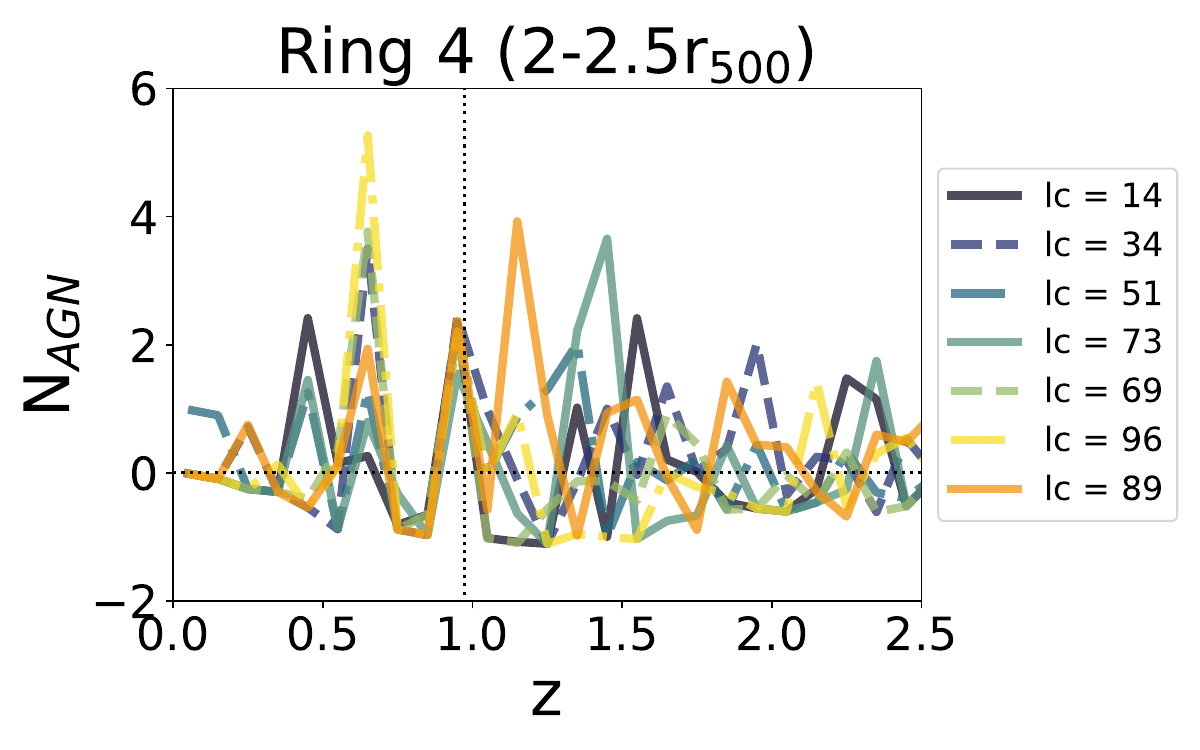}
    \caption{Redshift distribution of mock AGN that lie in the radial ring $r=2-2.5\,R_{500}$. Different colours correspond to each of the 7 randomly selected light-cone realisations of Figure \ref{fig:indiv_rad_distrib} (see legend) that reproduce an excess number of projected counts at the radial distance ring $r=2-2.5\,R_{500}$ in agreement with the observations presented in Figure \ref{fig:rad_distrib}.}
    \label{fig:z_distrib}
\end{figure}

For the simulated observations in Figure~\ref{fig:indiv_rad_distrib} that reproduce an excess of X-ray counts at $2.5\,R_{500}$ we further explore the redshift distribution of mock AGN. This is to investigate if the excess of X-ray sources is associated with the cluster. In Figure~\ref{fig:z_distrib} we show seven examples of the redshift distribution of mock AGN in the radial distance bin of $2.5\,R_{500}$. These are selected from light cone configurations in Figure~\ref{fig:indiv_rad_distrib} that reproduce an overdensity of mock AGN at $2.5\,R_{500}$. Most of these realisations show a redshift distribution where the peak is generated by objects in the foreground and/or the background of the cluster. There are also realisations where mock AGN that produce the overdensity are at redshifts similar to the cluster. We reiterate, however, that even in this case this is a projection effect because of the zero-order assumption of the model. Nevertheless, the contribution of these cases is marginal and most of the redshift distributions are dominated by foreground and/or background sources.

Next, we explore the incidence of excess projected X-ray counts in other $R_{500}$ rings around the simulated clusters. We adopt the same set of conditions defined above to identify in a quantitative manner excess counts. The only deviation is that the main ring within which overdensities are searched for varies between $r=1.5-3.5~R_{500}$. The model predicts an occurrence of about 10-20\% of an overdensity for different cluster centric distances. For distances smaller than  $r=1.5~R_{500}$ or bigger than  $r=3.5~R_{500}$, the sensitivity of the observation drops because of the extended emission of the cluster and the increasing off-axis incidence angles respectively. Hence, it is difficult to make a clear comparison at these radii. 

\begin{figure*}
    \centering 
    \includegraphics[width=.9\textwidth]{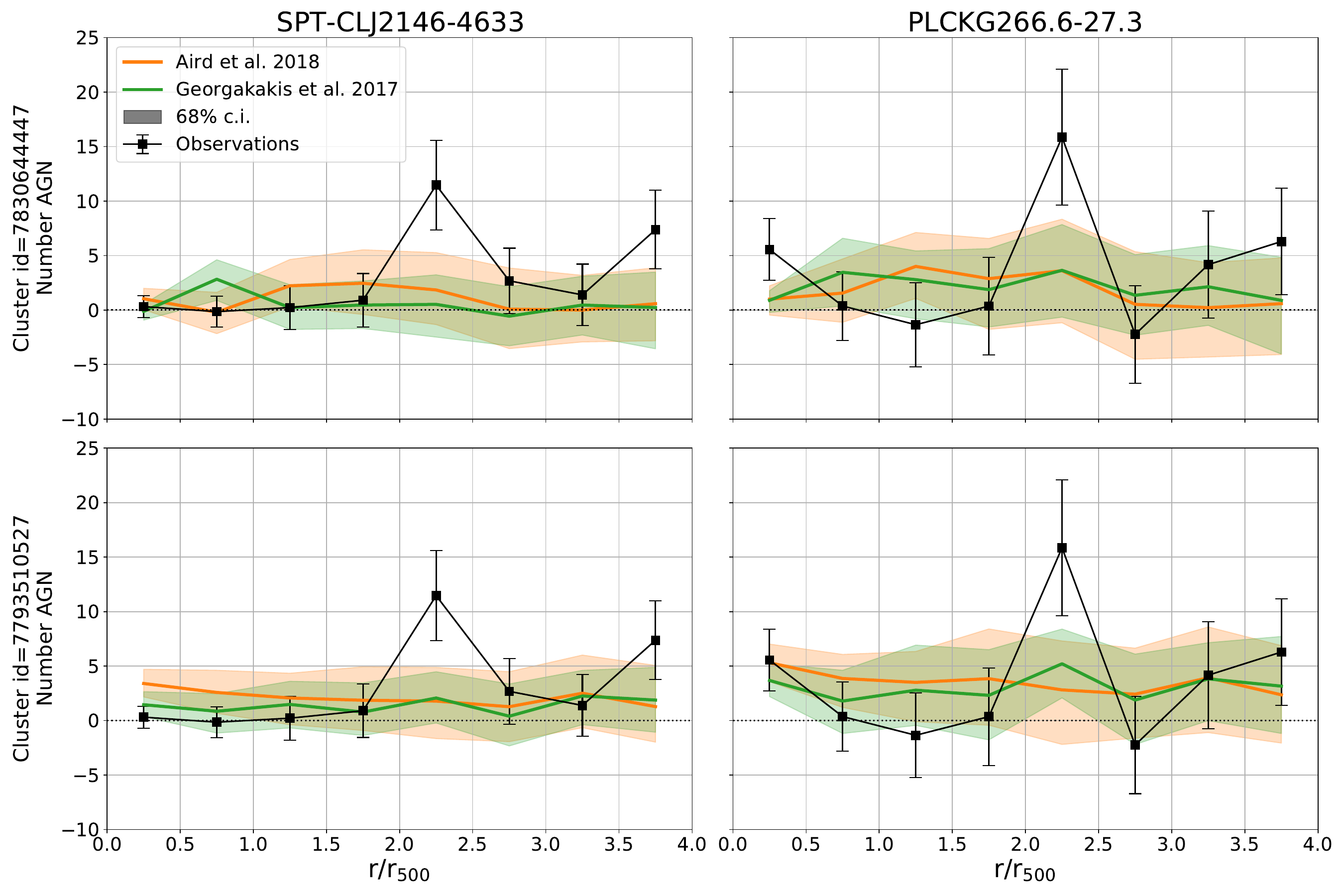}
    \caption{
    The observed projected radial distribution of X-ray selected AGN  (black squares and black solid connecting lines) of the clusters PLCKG266.6-27.3 (right column of panels) and SPT-CLJ2146-4633 (left column of panels) are compared with the semi-empirical model predictions (coloured lines and shaded regions). The two cluster observations differ in the total {\it Chandra} exposure time, with PLCKG266.6-27.3 being deeper (about 200\,ks) and  SPT-CLJ2146-4633 shallower (about 70\,ks). In all panels the green lines and shaded regions represent the model that uses the \citet{age17_sar} SARD, while the orange lines and shaded regions are for models that adopt the \citet{aird18_sar} SARD for seeding galaxies with AGN. The solid lines are the average of the 100 realisations, while the shaded regions indicate the $1\sigma$ scatter at fixed radial bin. The  model predictions are constructed for two different massive halos in \textsc{UniverseMachine}. The upper row of panels is for the halo with id=7830644447 and mass $M_{500c}\approx 7.5\times10^{14}$~M$_\odot/h$ (same as in Figures \ref{fig:rad_distrib}, \ref{fig:indiv_rad_distrib}). The lower row of panels is for the halo with id=7793510527 and mass $M_{500c}\approx 5\times10^{14}$~M$_\odot/h$ (see Figure \ref{fig:rad_distrib_obs}).  
    }
    \label{fig:rad_distrib_all_id_obs}
\end{figure*}

All the results above correspond to simulations of a single massive  halo (M$_{500c}\sim7.5\times 10^{14}$\,M$_\odot$) and the implementation of a single sensitivity map, the one that corresponds to the {\it Chandra} observations of PLCKG266.6-27.3 with a total on source exposure of $\approx 200$\,ks. Next we explore the impact of different X-ray depths in the result and conclusions. For this purpose we repeat the analysis using the same halo in the simulations to construct light cones but, applying a different sensitivity map to construct mock X-ray observations. The new map corresponds to the shallower {\it Chandra} observations of the cluster SPTCLJ2146-4633 with a total on source exposure of $\approx 70$\,ks. The main effect is that the number of detected AGN and the overall scatter decreases. This is because less luminous AGN are harder to detect in the case of shallower X-ray observations. Nevertheless, since the total number of AGN also decreases the fraction of mock observers that see an excess of projected X-ray counts at $r=2.5\,R_{500}$ based on the conditions presented earlier (see Equation~\ref{eq:overdensity_cond}) is similar to our baseline results using the most sensitive observation, i.e., about 10\% (see also upper panels of Figure~\ref{fig:rad_distrib_all_id_obs}). The effects of cluster mass onto the radial distribution are also investigated by repeating the same exercise for a different less massive halo in the N-body simulation with M$_{500c}\sim5\times 10^{14}$\,M$_\odot$ (\textsc{UniverseMachine} id=7793510527). The corresponding radial distributions for the different sensitivity maps (i.e. the ones of the observed clusters PLCKG266.6-27.3 and SPTCLJ2146-4633) are shown in the lower panels of Figure~\ref{fig:rad_distrib_all_id_obs}. In both cases we find a flat mean distribution with a large scatter around it which mimics our baseline result. This is an expected feature of the model since its zero-order assumption is that the AGN activation is independent of the halo mass.

\section{Discussion}\label{sec:discussion}
\subsection{Radial distribution of AGN}
The overarching question of this work {\bf relates to}  the role of the environment in modulating accretion events onto the SMBHs at the nuclear regions of galaxies. We approach this problem by investigating the X-ray AGN projected radial distribution in the vicinity of massive clusters of galaxies. These structures represent the densest regions in the Universe, where environmental effects and processes are expected to reach their maximum impact \citep[e.g., starvation, strangulation or ram-pressure, see][]{moore1996,larson1980,gunn1972}. It is now well established that the number of AGN in clusters of galaxies increases with redshift \citep[e.g.][]{martini09, martini13}. This trend mirrors the evolution of the overall AGN field population \citep[e.g.][]{ueda2014,aird2015} and perhaps proceeds even faster \citep[e.g.][]{Ehlert14,bufanda17,Hashiguchi2023,Toba2024}, thereby suggesting that dense environments at high redshift promote accretion events onto SMBHs \citep[e.g.][]{lehmer09,digby-north10}. It has been proposed that the incidence of AGN in massive clusters is related to an infalling population of galaxies whose black holes become active as they enter the dense cluster environment \citep[e.g.][]{haines12,pimbblet13}.

In this work we test this scenario by modelling the observed projected radial distribution of X-ray selected AGN in massive clusters at $z\approx1$ presented by \cite{koulouridis2019}. That cluster sample is advantageous because the individual cluster properties (mass and radius) are accurately determined using a sophisticated method that combines information from both \textit{XMM-Newton} and \textit{Chandra} observations \citep[see][]{Bartalucci2017,Bartalucci2018}. The large effective area of the former allows the characterisation of faint structures, while the spatial resolution of the latter enables  modelling the central regions of the clusters. This approach leads to an accurate characterisation of the density profile of the clusters out to $R_{500}$. For this sample it is therefore possible to build robust radial distributions of X-ray selected AGN as function of distance normalised to $R_{500}$ and explore evidence for a statistically significant excess of counts at the radius $2.5\,R_{500}$.

The semi-empirical modelling developed in this work emphasises the role of sample variance in the interpretation of the observed projected AGN radial distributions. We produce mock AGN catalogues under the explicit assumption that accretion events on the SMBHs are triggered with the same probability in the different environments. Then we use these mock AGN and galaxy catalogues to simulate realistic observations of clusters that include the same selection effects as the observations of \cite{koulouridis2019}. We study the impact of projection effects by simulating 100 observations of the same cluster in the simulation with randomly selected lines of sight (see Section~\ref{sec:impl_ths_wrk}). 
A striking results from our analysis is the flatness of the simulated average projected radial distribution (see Figure~\ref{fig:rad_distrib}), which at first instance appears inconsistent with the observations of \cite{koulouridis2019}. At the same time however, there is substantial scatter around the mean of this distribution as a result of sample variance, i.e. background/foreground structures along the line of sight projecting into the field of view (see Fig. \ref{fig:z_distrib}). Given this scatter the significance of the excess counts at the radial ring $2.5\,R_{500}$ in Figure~\ref{fig:rad_distrib} is significant only at the $1-2\,\sigma$ level. We nevertheless, take a further step and calculate the probability of finding overdensities similar to the observed ones. This  analysis also demonstrates the importance of stochasticity in producing excess X-ray AGN counts in the radial distribution of counts that have no physical origin. The model reproduces radial distribution overdensities at $2.5\,R_{500}$ similar to those found by \cite{koulouridis2019} in up to 20\% of the simulated light cones (see Figure~\ref{fig:indiv_rad_distrib}). This fraction should be compared with the rate of 40$\pm$20\% (2 out of a total of 5 clusters) in the sample of \cite{koulouridis2019} that show a statistically significant excess of counts. These results also have implications for other studies in the literature that find evidence for excess AGN counts in the projected radial distribution of AGN beyond the virial radius \citep{Johnson03, ruderman05,fassbender12}. 

We caution that our simulations cannot {\bf reject the possibility of} a physical interpretation of the excess counts at $2.5\,R_{500}$ found by \cite{koulouridis2019}. Addressing the origin of this excess, physical or stochastic, requires spectroscopic information, which would allow the robust identification of AGN cluster members and separate them from foreground/background interlopers. Increasing the cluster sample will allow a better understanding of the physics at play. This is because different studies show that the dynamical state (i.e., relaxed vs. non-relaxed) of the cluster could have an impact on the AGN activity \citep[see][]{Kocevski09,vanBreukelen09,stroe21} and the two clusters which show the overdensity in \cite{koulouridis2019} are in different relaxation states, i.e., one of them is virialised while the other is not \citep[see][]{Bartalucci2017}. 

\subsection{Exploring a higher incidence of AGN among the infalling mock galaxy population}
Next, we test the hypothesis of an infalling population as the origin of the excess counts in the radial distribution of AGN at about $2\,R_{500}$ in Figure~\ref{fig:rad_distrib}.  Our approach is to  tune our semi-empirical model by associating a higher incidence of AGN among infalling galaxies. This requires (i) a criterion for isolating galaxies that enter for the first time the cluster from the cosmic web and (ii) a new specific accretion rate distribution model that is applied to these galaxies and has the property of producing a higher incidence of AGN at fixed X-ray luminosity threshold.

\begin{figure}
    \centering 
    \includegraphics[width=.5\textwidth]{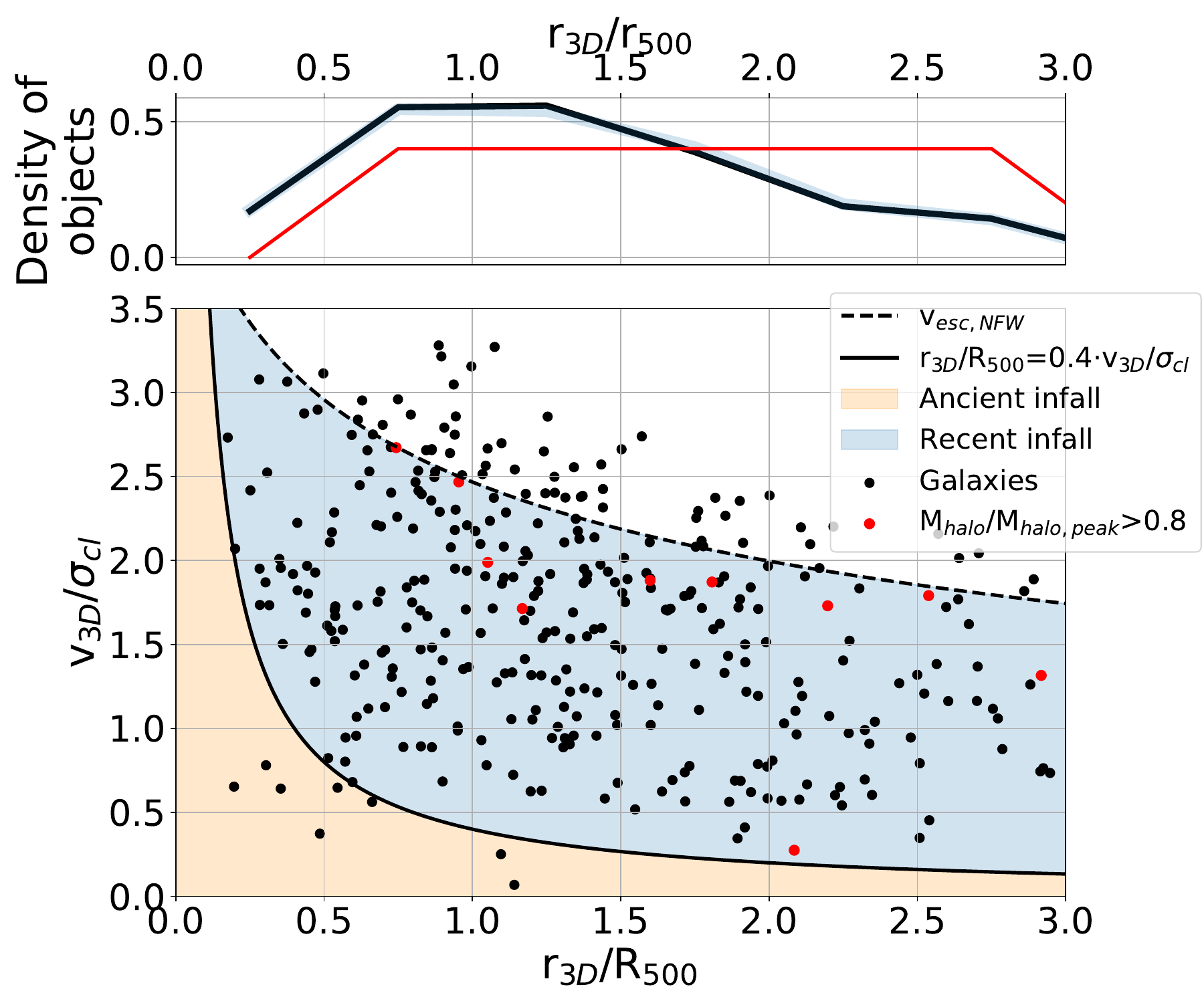}
    \caption{The 3-dimensional phase-space diagram used to identify the infalling galaxy population of the cluster with id=7830644447 in \textsc{UniverseMachine}. Black dots correspond to individual galaxies in \textsc{UniverseMachine}. The blue shaded area marks the recent infall region of the parameter space and is defined by the caustic $\frac{r_{3D}/R_{500}}{v_{3D}/\sigma_{cl}}=0.4$ \citep[black solid line, e.g.,][]{2023Kim} and the escape velocity of the equivalent NFW halo profile (dashed black line). The orange shaded area under the caustic $\frac{r_{3D}/R_{500}}{v_{3D}/\sigma_{cl}}=0.4$ is often referred to as ancient infall or first infallers region of the phase-space diagram. The red dots represent recent infall galaxies with dark matter halo masses that have at least 80\% of their maximum historical masses ($\rm{M}_{halo,peak}$ parameter in \textsc{UniverseMachine} catalogue). These are the halos that we consider as infalling in our analysis. The panel at the top shows the (normalised) radial distribution histogram of the different galaxy populations with the same color coding, black refers to the whole population of galaxies, blue to the galaxies in the recent infall region (i.e. those within the blue shaded area) and red for the infall galaxies which dark matter haloes have at least 80\% of their maximum historical masses.}
    \label{fig:phase-space}
\end{figure}

Ideally, an infall population would be defined by following the orbits of the dark matter particles that make up halos. However, semi-empirical models, like \textsc{UniverseMachine}, are build upon halo catalogues and therefore information about the formation/merging  history of individual halos is not readily available. Instead, we decide to adopt the alternative but widely used approach of the phase-space diagram to find infalling halos. For a given massive cluster halo in \textsc{UniverseMachine} it is possible to estimate the relative velocities ($v_{3D}$) and relative radial distances ($r_{3D}$) to other halos in the simulation (parent or satellites). The phase-space diagram of the cluster under consideration is then defined by the parameters  $v_{3D}/\sigma_{cl}$ ($\sigma_{cl}$ is the velocity dispersion of the main cluster halo) and   $r_{3D}/R_{500}$ ($R_{500}$ refers to the main cluster halo). Halos with different infall histories populate distinct regions of the phase-space plane. This is  because the ratio between $r_{3D}/R_{500}$ and $v_{3D}/\sigma_{cl}$ is a proxy of the infall time of a main cluster halo \citep[see e.g.,][]{2013Noble,2016Noble,2017Rhee,2023Kim}. In the notation above the $3D$ index refers to 3-dimensional quantities estimated from the spatial distribution of halos in the \textsc{UniverseMachine} simulation box. The adopted $3D$ phase-space diagram is independent of projection effects that are inevitable when constructing light cone realisations assuming different observer positions (see Section \ref{sec:light-cones}).  Following commonly used criteria we define infalling halos/galaxies as those or that simultaneously satisfy the following conditions

\begin{equation}
    \begin{split}
        & r_{3D}<3~R_{500}\\
        & |\Delta v|<3.5\sigma_{cl}\\
        & v_{3D} < v_{esc,NFW}\\
        & \frac{r_{3D}/R_{500}}{v_{3D}/\sigma_{cl}} > 0.4\\
        & \rm{M}_{halo}/\rm{M}_{halo,peak}>0.8.   
    \end{split}
    \label{eq:infall}
\end{equation}

\noindent where $v_{esc,\rm{NFW}}$ corresponds to the escape velocity \citep[e.g.][]{2017Rhee} of a halo assuming an Navarro-Frenk-White profile \citep[NFW,][]{1996Navarro} and $\rm{M}_{halo,peak}$ is the maximum historical mass of the halo. Figure~\ref{fig:phase-space} shows the phase space diagram for the cluster with dark matter halo id=7830644447, i.e. the same massive halo used to construct light cones and simulated radial distributions (see Section~\ref{sec:impl_ths_wrk}). The sample of infalling galaxies based on the conditions above is indicated with the red circles in Figure~\ref{fig:phase-space}. 

\begin{figure}
    \centering 
    \includegraphics[width=.5\textwidth]{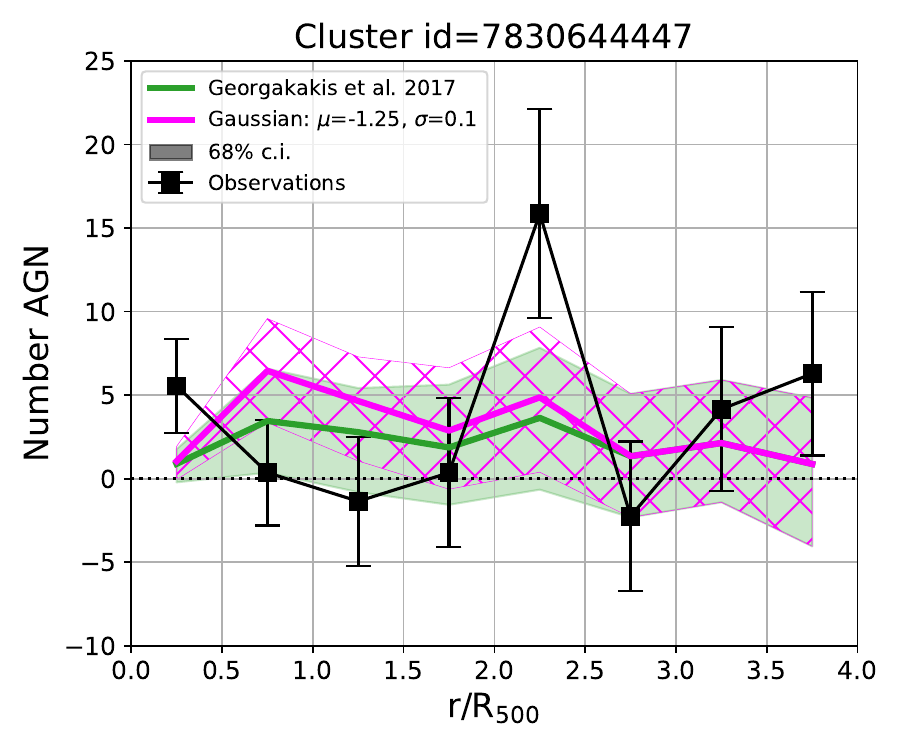}
    \caption{The X-ray AGN radial distribution. The observations (black squares and sold lines) and the semi-empirical model predictions (lines and shaded regions) are plotted at different cluster centric distances normalized to $R_{500}$. Black points connected with the solid black line represent the observed radial distribution of the cluster PLCKG266.6-27.3. The solid green line is the mean semi-empirical model prediction in the case of the \citet{age17_sar} specific accretion rate distribution. The magenta solid line correspond to the semi-empirical model in which the modified SARD described in Section \ref{sec:SEM} is applied to the infalling galaxy population identified in Figure \ref{fig:phase-space}.  The light-green shaded and magenta hatched regions within which the semi-empirical model lines are embedded correspond to the 68\% confidence intervals of the mean value. They represent the variance between different lines of sight (see the text for further details). Both semi-empirical model predictions are for the massive halo with id=7830644447 in \textsc{UniverseMachine} with virial mass $\sim 8.1\cdot10^{14}$~M$_\odot/h$.}
    \label{fig:rad_distrib_infall}
\end{figure}

The next step is to adopt a new specific accretion rate distribution model, which when applied to the infalling galaxies above yields a higher fraction of AGN. In \cite{munozRodriguez2023} we showed that the observed fraction of X-ray selected AGN relative to galaxies in massive clusters of galaxies at $z\approx1$ is much higher than that predicted by our baseline semi-empirical model that uses either the \cite{age17_sar} or the \cite{aird18_sar} SARs. Instead, \cite{munozRodriguez2023} proposed that a log-normal SAR model with mean specific accretion rate $\log \lambda_{SAR}=-1.25$ and scatter $\sigma=0.1$ applied to galaxies with stellar masses M$_*>10^{10.7}$~M$_\odot$ can reconcile the tension with the observed fraction of X-ray selected AGN in massive clusters at $z\approx1$. We therefore choose to use the same SAR model in our analysis and apply it only to the infalling galaxies (red circles) of Figure~\ref{fig:phase-space}. The impact on the AGN radial distribution of the modified SAR for the infalling galaxies is shown in Figure~\ref{fig:rad_distrib_infall}. Relative to our baseline model the mean expected number of X-ray selected AGN slightly increases for the radial ring $2-2.5\,R_{500}$, i.e. the one where excess counts where observed by \cite{koulouridis2019}. However, the same effect is seen at smaller radii, $0.5-2\,R_{500}$. This is because the infall population is evenly distributed between $r_{3D}=0.5-3\,R_{500}$ as demonstrated by the top panel of Figure~\ref{fig:phase-space}. In any case, the increase at the ring $2-2.5\,R_{500}$ is modest and is associated with substantial scatter. We apply the criteria of equation \ref{eq:overdensity_cond} to identify in an objective manner excess counts in the ring $2-2.5\,R_{500}$ among the light-cone realisations with the modified SAR. We find that 20\% of the light cones show radial distributions that resemble the observations. This rate is the same as with the baseline semi-empirical model  predictions presented in Figure \ref{fig:indiv_rad_distrib}. We conclude that the approach outlined above for increasing the incidence of AGN among infalling galaxies in massive clusters has a moderate impact on the observed radial distribution of AGN and cannot fundamentally modify the predictions of our baseline semi-empirical model. 

\section{Conclusions}

In this paper we develop a flexible semi-empirical model of AGN and galaxies in a cosmological volume to interpret observations of the radial distribution of AGN in massive clusters of galaxies at $z\approx1$ \citep{koulouridis2019} and test claims for an efficient activation of SMBHs in the outskirts of galaxy clusters. The explicit assumption of the model is that the AGN triggering is independent of environment (or halo mass). This allows us to test the hypothesis that the excess counts of X-ray selected AGN observed at a radius of about $2-2.5\, R_{500}$ in massive clusters of galaxies at $z\approx1$ \citep{koulouridis2019} are not physical but instead driven by projection effects.

We select halos at $z\approx1$ in the simulations with masses similar to the clusters of \cite{koulouridis2019} and generate mock observations through different sight lines to test the impact of sample variance to the inferred mock AGN radial distribution. A key step of this process is the generation of light cones which allows us to implement the selection effects of the real observations to the mocks (e.g. field-of-view, variations of the flux limit at different radial distances from the cluster centre). The main results of the paper are:

\begin{enumerate}
    \item We demonstrate that our semi-empirical model predicts HODs for X-ray selected AGN in broad agreement with the latest observational constraints of \cite{2023Comparat} at $z\sim0.2$. The normalisation of our HODs decreases toward lower redshift and brighter luminosities, mirroring the evolution of the X-ray AGN population with redshift and the form of the X-ray luminosity function. 

    \item There is evidence for a possible tension between observations and model predictions on the HOD slope of satellite AGN. The observations favour flatter slopes compared to the semi-empirical model. Although the observational uncertainties are large, this discrepancy may point to the suppression of X-ray AGN  in satellites galaxies of massive cluster of galaxies at $z\approx0.25$.
    
    \item Turning to the projected radial distribution of X-ray selected AGN in the vicinity of massive clusters at $z\approx1$, our model predicts on average a flat radial distribution. This is a direct consequence of the main assumption of the model construction that the AGN triggering is independent of the environment (Figure~\ref{fig:rad_distrib}). 
 
    \item Our analysis emphasises the importance of sample variance that manifests as scatter around the mean of the projected radial distributions predicted by the model. As a result in a non-negligible number of cases excess counts at radial distances of 2--2.5$\,R_{500}$ are predicted by the model. Up to 20\% of the realisations show amplitudes similar to the observations of \cite{koulouridis2019} for massive clusters of galaxies at $z\approx1$ (see Figure~\ref{fig:indiv_rad_distrib}). This incidence rate is lower but still consistent within the errors with the observed fraction of clusters in the \cite{koulouridis2019} work with excess counts in their outskirts, $40\pm20$\%. In our model, however, these overdensities in the projected radial distribution are not physical but stochastic and dominated by interlopers (Figure~\ref{fig:z_distrib}).

    \item Fine tuning our model to favour a higher incidence of mock AGN among galaxies in the infall region of masssive halos has little impact to the predicted projected radial distributions (see Figure~\ref{fig:rad_distrib_infall}). This further emphasises the significance of sample variance in interpreting projected AGN radial distributions.   
\end{enumerate}

\section*{Acknowledgements} 
This project has received funding from the European Union’s Horizon 2020 research and innovation program under the Marie Skłodowska-Curie grant agreement No 860744. AG and IMR acknowledge support from the Hellenic Foundation for Research and Innovation (HFRI) project "4MOVE-U" grant agreement 2688, which is part of the programme "2nd Call for HFRI Research Projects to support Faculty Members and Researchers". AL is partly supported by: "Data Science methods for MultiMessenger Astrophysics \& Multi-Survey Cosmology", Programmazione triennale 2021/2023 (DM n.2503 dd. 9 December 2019), Programma Congiunto Scuole; Fondazione ICSC, Spoke 3 Astrophysics and Cosmos Observations  Project ID CN-00000013. This research made use of Astropy,\footnote{http://www.astropy.org} a community-developed core Python package for Astronomy \citep{astropy:2013, astropy:2018}; \texttt{Colossus}\footnote{https://bdiemer.bitbucket.io/colossus/index.html} \citep{diemer18_colossus}, \texttt{Numpy}\footnote{https://numpy.org/} \citep{numpy2011}, \texttt{Scipy}\footnote{https://scipy.org/} \citep{2020SciPy-NMeth}, \texttt{Matplotlib}\footnote{https://matplotlib.org/} \citep{matplotlib2007}, NASA’s Astrophysics Data System (ADS) and the arXiv preprint server.

\section*{Data Availability}
The dataset of 100 simulated light-cones used in the analysis will be made available in Zenodo upon acceptance of the paper. The light-cone generation code is available upon request and will be public on github in the near future.


\bibliographystyle{mnras}
\bibliography{biblio} 






\bsp	
\label{lastpage}
\end{document}